\documentclass[pra,twocolumn,superscriptaddress]{revtex4}
\usepackage{graphicx}
\usepackage{amsmath}
\usepackage{mathtools}
\usepackage{cases}
\usepackage{graphicx}
\usepackage{float}

\renewcommand{\thefigure}{\arabic{figure}}

\renewcommand{\thetable}{\arabic{table}}

\begin{document}
 
 \newcommand{\beginsupplement}{%
        \setcounter{table}{0}
        \renewcommand{\thetable}{S\arabic{table}}%
        \setcounter{figure}{0}
        \renewcommand{\thefigure}{S\arabic{figure}}%
     }
 
\title{Energy-dependent quenching adjusts the excitation diffusion length to regulate photosynthetic light harvesting}

\author{Doran I. G. Bennett}
\thanks{These authors contributed equally to this work.}
\affiliation{Department of Chemistry and Chemical Biology, Harvard University, Cambridge, MA 02138}
\author{Graham R. Fleming}
\affiliation{Department of Chemistry, University of California; Molecular Biophysics and Integrated Bioimaging Division, Lawrence Berkeley National Labs, Berkeley, CA 94720}
\author{Kapil Amarnath}
\thanks{These authors contributed equally to this work.}
\affiliation{FAS Center for Systems Biology, Harvard University, Cambridge, MA 02138}

\begin{abstract}
An important determinant of crop yields is the regulation of photosystem II (PSII) light harvesting by energy-dependent quenching (qE). However, the molecular details of excitation quenching have not been quantitatively connected to the PSII yield, which only emerges on the 100 nm scale of the grana membrane and determines flux to downstream metabolism. Here, we incorporate excitation dissipation by qE into a pigment-scale model of excitation transfer and trapping for a 200 nm $\times$ 200 nm patch of the grana membrane. We demonstrate that single molecule measurements of qE are consistent with a weak-quenching regime. Consequently, excitation transport can be rigorously coarse-grained to a 2D random walk with an excitation diffusion length determined by the extent of quenching. A diffusion-corrected lake model substantially improves the PSII yield determined from variable chlorophyll fluorescence measurements and offers an improved model of PSII for photosynthetic metabolism.
\end{abstract}

\maketitle

Plants fix $\sim$60 petagrams of carbon every year \cite{Field1998} and are an essential food source.  Roughly two-thirds of harvested global crop calories come from four crops \cite{Long2015} and optimizing yields to feed the growing population is an important goal.  Predicting how photosynthetic metabolism and crop yield change in response to genetic and environmental perturbations constitutes a grand challenge for science.  Though much has been learned about the molecular mechanisms of light harvesting and charge separation,  using this knowledge to create a tractable but rigorously defined multiscale model at the membrane level remains a complex challenge.  In this paper we build on our earlier models of energy transport in photosystem II of plants \cite{Bennett2013,Amarnath2016} to create a model of the rapidly reversible portion of plants' response to excess light in a 200 nm $\times$ 200 nm patch of the grana membrane.  In doing so we are able to identify a single variable, the excitation diffusion length, which controls the response of plant light harvesting systems to rapid changes in light level.  This regulatory system is important for plant fitness \cite{Kulheim2002} and crop yield \cite{Kromdijk2016}.

\begin{figure}[t!]
 \begin{center}
 \includegraphics{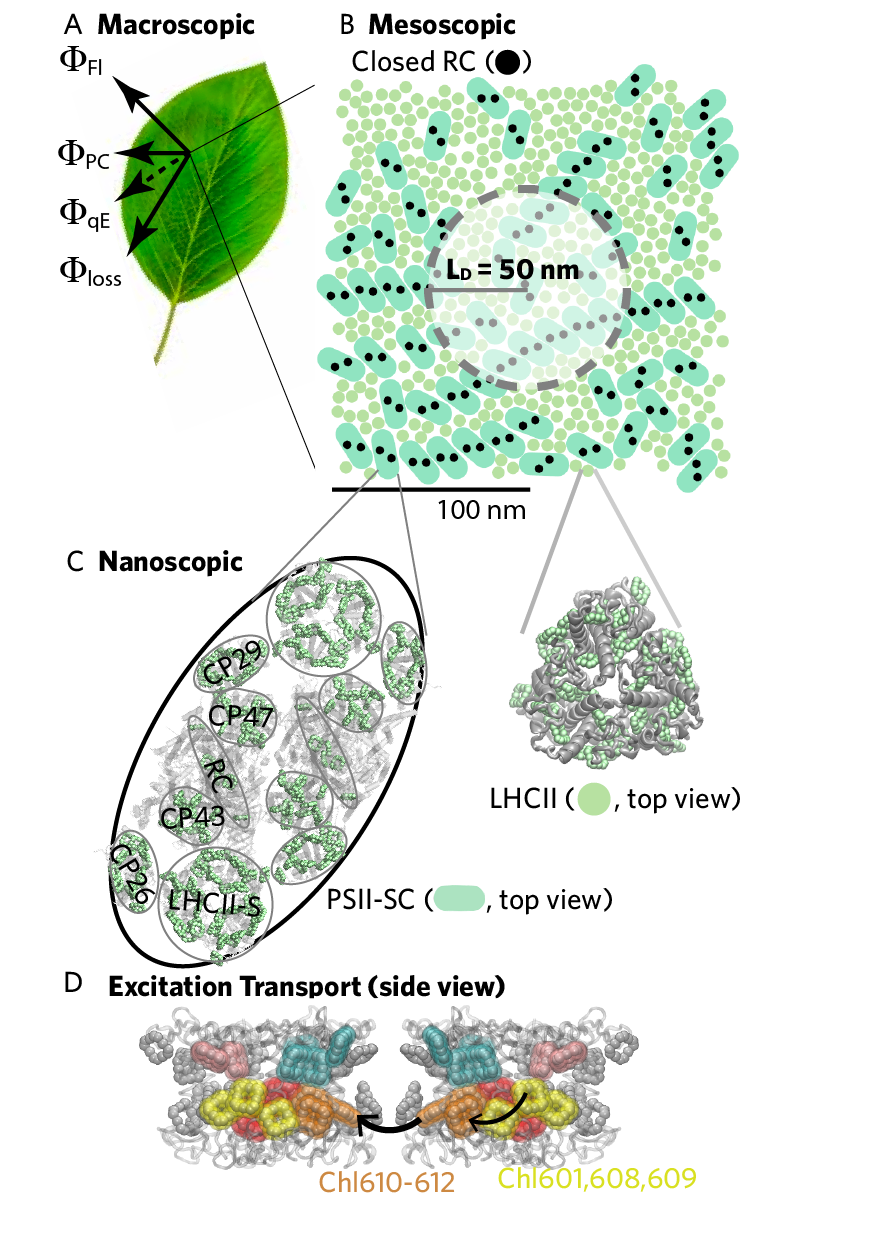}
 \caption{\textbf{The multiple scales of PSII light harvesting.} \textbf{(A)} The photosynthetic output of a leaf is affected by the quantum yields of the different decay pathways for excitation in PSII: productive photochemistry in the reaction centers, $\Phi_{\textrm{PC}}$; emission as fluorescence, $\Phi_{\textrm{Fl}}$; loss due to non-radiative decay, $\Phi_{\textrm{loss}}$; and dissipation by activated quenching sites, $\Phi_{\textrm{qE}}$. \textbf{(B)} Photosystem II harvests sunlight across the mesoscopic ($\sim$100s of nm) thylakoid membrane. It consists of photosystem II supercomplexes (pills) and major light harvesting complex II trimers (LHCII, circles). The filled black circles indicate closed reaction centers. The shaded circle with a radius equivalent to the excitation diffusion length ($L_{\textrm{D}}$ = 50 nm) indicates the spatial extent of excitation transport from an initial excitation at the center of the circle. \textbf{(C)} The multiscale model is composed of the crystal structures for PSII supercomplexes \cite{Caffarri2009} and unbound LHCII \cite{Liu2004}. The PSII supercomplex is a dimer, with each monomer containing one RC, and a pair of core antenna proteins (CP43 and CP47), a pair of minor LHCs (CP26 and CP29) replaced by LHCII monomers, and a strongly bound LHCII (LHCII-S). The pigments are indicated in light green, and the surrounding protein scaffold in gray. \textbf{(D)} Energy transfer (black arrows) is described using Generalized F\"{o}rster theory between domains of $\sim$3-4 tightly coupled chlorophylls (colored pigments).  \label{fig1}}
 \end{center}
 \end{figure}

Nonphotochemical quenching regulates photosystem II (PSII) light harvesting by dissipating excess absorbed sunlight in the pigment-protein complexes that serve as antenna. In dim sunlight a photon absorbed by an antenna complex results in a nascent excitation that is efficiently delivered to a reaction center (RC), where charge separation converts the excitation energy to chemical energy. The fraction of excitations that result in productive charge separation ($\Phi_{\textrm{PC}}$, photochemical yield, Fig.~\ref{fig1}A) is $\sim$83\% in optimal conditions \cite{Baker2008}. A brief period of intense light, or sunfleck \cite{Way2012}, results in a transient increase in the flux of photochemistry at the reaction centers. Consequently the pH gradient across the thylakoid membrane increases, and the fraction of open reaction centers available for performing charge separation decreases. In response to the increased pH gradient, the largest, rapidly reversible component of NPQ, qE (\underline{e}nergy-dependent \underline{q}uenching), activates, and specific pigment sites dissipate excitation energy in the antenna. By decreasing the flux of excitation that reaches the reaction centers, qE increases the fraction of open RCs and decreases the fraction of excitations that damage closed reaction centers momentarily occupied with charge separation \cite{Kulheim2002,Li2002}. In this fashion qE optimizes the balance between the energetic benefit of photochemistry and the metabolic cost of reaction center damage, while meeting the demands of downstream reactions such as CO$_2$ fixation \cite{Genty1989,Kanazawa2002}. Quantifying the net metabolic benefit of qE thus requires an accurate description of how it influences the photochemical yield ($\Phi_{\textrm{PC}}$).

The key challenge to establishing a quantitative relationship between qE and the photochemical yield ($\Phi_{\textrm{PC}}$) is reconciling events occurring on the nano- and mesoscale. While qE acts on the pigment scale (Fig.~\ref{fig1}C), the photochemical yield is the result of all productive charge separation events occurring at open reaction centers across the thylakoid membrane (Fig.~\ref{fig1}A-B). Looking from the nanoscale up, several different pigment sites and photophysical mechanisms of quenching in the antenna complexes have been proposed \cite{Duffy2015}, and it remains unclear how these details influence the photochemical yield. From the top down the photochemical yield is determined by applying phenomenological (`lake' and `puddle') models \cite{Robinson1966} to the chlorophyll fluorescence yield (Fig.~\ref{fig1}A,  $\Phi_{\textrm{Fl}}$) of a leaf exposed to fluctuating light \cite{Kramer2004,Baker2008}. However, it remains unclear how well these models describe the interaction between qE and PSII light harvesting because they neglect any of the molecular details of quenching as well as the excitation transport that occurs across tens of nanometers in the thylakoid membrane \cite{Bennett2013,Amarnath2016}. Thus, a multiscale model is required to explicitly calculate the photochemical yield from the light harvesting dynamics of a PSII membrane containing $\sim$50,000 pigments in the presence of qE. Such a model could reveal simplifying principles at the mesoscale that connect qE to the photochemical yield. 

In this paper, using a pigment scale model of excitation energy transfer and quenching, we show that the excitation diffusion length is  the key degree of freedom connecting the molecular mechanism of qE quenching to PSII yields. Using available single-molecule spectroscopic data, we show that qE occurs in the weak-quenching regime. As a result, qE simply modulates the extent of 2D diffusion via the excitation diffusion length to control how much excitation reaches the reaction centers. The commonly used lake and the puddle models inadequately describe the influence of qE on the photochemical yield of PSII because they assume a constant excitation diffusion length. However a straightforward diffusion correction to the lake model substantially improves its ability to relate the excitation diffusion length and the fraction of open reaction centers to the photochemical yield. We believe that this rigorous coarse-graining of PSII light harvesting will prove useful for quantitatively connecting the molecular mechanisms of individual complexes to their functional role in photosynthetic metabolism and growth. 

\section*{\large{Results}}
\subsection{Emergence of 2D diffusive excitation transport.}
The modeling of ultrafast spectroscopic data of pigment-protein complexes isolated from the thylakoid membrane has revealed a great deal about the mechanisms of excitation energy transport between pigments \cite{Ishizaki2010}. However, the functional role of antenna proteins, to deliver excitations to reaction centers, only emerges on the length scale of the intact thylakoid membrane ($\sim$100 nm). {\it In vivo}, a nascent excitation transfers across the PSII-enriched portion of the thylakoid (PSII) membrane (Mesoscopic, Fig.~\ref{fig1}B) through a dense network of protein-bound chlorophylls predominately associated with the major light harvesting complex II trimers (LHCII, Fig.~\ref{fig1}C) and PSII supercomplexes (Fig.~\ref{fig1}C) until it reaches the reaction center where charge separation takes place. Therefore to understand the connection between excitation transfer in antenna proteins and charge separation at reaction centers requires understanding the emergent properties that characterize excitation dynamics at the mesoscale.

\begin{figure}[t!]
\begin{center}
\includegraphics[scale=0.227]{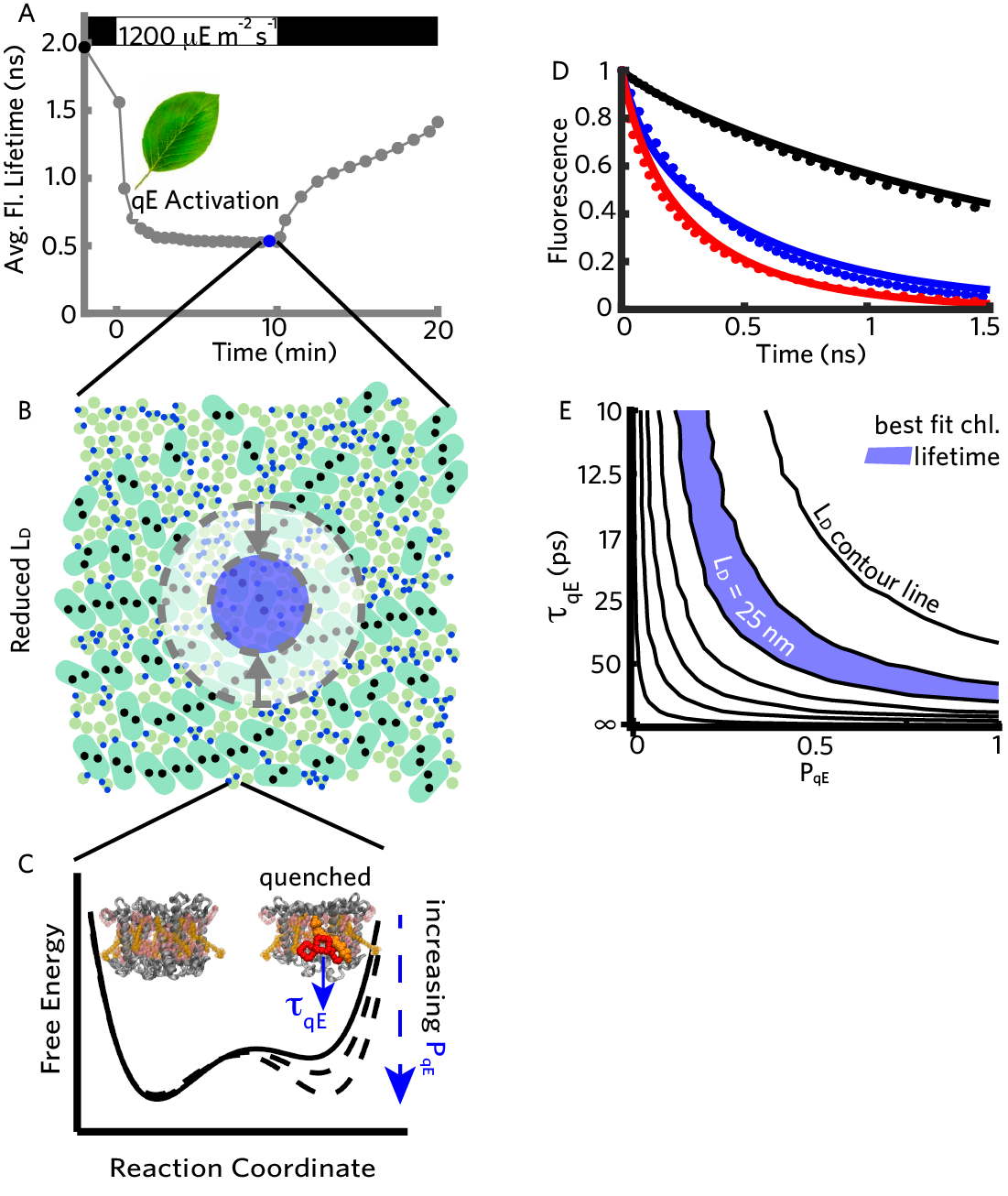}
\caption{\textbf{The diffusion length defines the excitation dynamics in the presence of weak quenchers.} \textbf{(A)} The average fluorescence lifetime when all reaction centers are closed of a wild type \emph{A. thaliana} leaf exposed to a dark-light-dark sequence. The black and blue dots represent the correspondingly colored fluorescence lifetime decays (solid lines) in D. \textbf{(B)} The PSII membrane in a representative qE configuration underlying the corresponding measurement in (A). The filled black circles indicate closed reaction centers while the blue dots indicate activated quenching sites using the LHCII 610-612 domain. The reduction of excitation diffusion length ($L_{\textrm{D}}$) by quenching sites is shown. \textbf{(C)} Model for quenching. An antenna complex monomer is treated as a two-state switch, with a probability $P_{\textrm{qE}}$ of the quenching site (domain) having a time scale of quenching $\tau_{\textrm{qE}}$. \textbf{(D)} Simulation (dotted lines) of fluorescence lifetime measurements (solid lines) taken on intact membranes or leaves in different states. Red indicates a state of open reaction centers with no qE, black indicates closed reaction centers with no qE, and blue indicates closed reaction centers with qE induced by exposure for 6.5 min to 1200 $\mu$mol photons m$^{-2}$ s$^{-1}$ light. Open RC data is from ref. \cite{vanOort2010} and closed RC data from ref. \cite{Sylak-Glassman2016}. \textbf{(E)} Contour plot of excitation diffusion length ($L_\textrm{D}$) as a function of $\tau_{\textrm{qE}}$ and $P_{\textrm{qE}}$. The blue area indicates an $L_\textrm{D}$ of $25\pm 2.5$ nm, in agreement with the best fit chlorophyll fluorescence lifetime measurements. \newline \newline \label{fig2}}
\end{center}
\end{figure}

While there are currently no spectroscopic techniques for characterizing excitation transport in PSII on the 100 nm length scale, we have previously combined structural, spectroscopic, and biochemical data to build a multiscale model of PSII light harvesting \cite{Amarnath2016, Bennett2013}. The model treats excitation transport between domains of $\sim$3-4 tightly coupled chlorophylls using Generalized F{\"o}rster theory \cite{Bennett2013,Sumi1999,Scholes2000} (Fig.~\ref{fig1}C, Excitation Transport, domains indicated by colors), with Hamiltonian parameters extracted from spectroscopic measurements of isolated pigment protein complexes and the interaction of pigments in different proteins treated as dipole-dipole coupling. Generalized F{\"o}rster provides the most coarse-grained model of energy transfer that correctly reproduces the dynamics from more quantum-mechanically exact simulations \cite{Roden2016,Kreisbeck2014}. With a simple kinetic model for charge separation parametrized on data of isolated PSII supercomplexes with variously sized antenna \cite{Bennett2013} and a method to get positions for LHCIIs and PSII supercomplexes in a 200 nm $\times$ 200 nm patch of the PSII membrane \cite{Schneider2013}, we simulated data taken on an intact membrane. Our model reproduces  \cite{Amarnath2016}, in the absence of free parameters, (i) the photochemical yield of dark acclimated leaves as measured by chlorophyll fluorescence yield, (ii) the chlorophyll fluorescence lifetime measured on intact membranes when all reaction centers are open (Fig.~\ref{fig2}D, red), and (iii) the hyperbolic shape of oxygen evolution as a function of the fraction of open reaction centers as measured originally by Joliot and Joliot in 1964 \cite{Joliot1964}. 

In the model, following an initial excitation, the variance of the population distribution increases linearly in time, which supports two-dimensional diffusive transport as an emergent property on the mesoscale \cite{Amarnath2016}. This effective dimensionality agrees with an alternative approach which fit fluorescence data to a fractional dimensional ($d$) random walk and found a value of approximately 2 ($d$ = 1.9 for intact thylakoid membranes and 2.2 for BBY preparations) \cite{Chmeliov2016a}. The diffusive transport is characterized by an excitation diffusion length through the antenna pigments ($L_{\textrm{D}}$, Fig.~\ref{fig1}B) of 50 nm \cite{Amarnath2016}. $L_{\textrm{D}}$ is defined here as the minimum net displacement in one dimension achieved by 37\% of the excitation population when all reaction centers are closed. The high quantum efficiency of PSII in dim light condition ($>$80\%) and the hyperbolic shape of oxygen evolution as a function of the fraction of open reaction centers are explained by the large number of reaction centers within 50 nm radius of an initial excitation (Fig.~\ref{fig1}B).

\subsection{qE controls the excitation diffusion length.}
As an \emph{A. thaliana} leaf acclimates to a sudden exposure to bright light (Fig.~\ref{fig2}A, 1200 $\mu$mol photons m$^{-2}$ s$^{-1}$), the average chlorophyll fluorescence lifetime when all RCs are closed decreases from $\sim$2000 ps to $\sim$550 ps over the course of 10 minutes (Fig.~\ref{fig2}A and D). Underlying this chlorophyll fluorescence quenching is the activation of numerous quenching sites across the membrane (Fig.~\ref{fig2}B, blue dots), a process which is collectively known as {\underline e}nergy-dependent {\underline q}uenching (qE). The activation of qE is the result of a conformational change in antenna proteins (Fig.~\ref{fig2}C) in response to a growing pH gradient across the PSII membrane. Since both the photophysical mechanism of quenching and the identity of the pigments that constitute the quenching site remain controversial \cite{Duffy2015}, it has been difficult to establish how qE influences excitation transport and light harvesting in PSII. 

In dark-acclimated leaves our model predicts 2D diffusive transport, but the kinetic competition between qE quenching and transfer out of an activated site could change the emergent excitation dynamics. In particular, if the rate of quenching greatly exceeds the rate of transfer out of the quenching site then excitation density will locally deplete and the overall process of transport can no longer be completely described by 2D diffusion. We will refer to this scenario as the `€™strong quenching'€™ regime. On the other hand, if the rate of quenching is much slower than the rate of transport out of the quenching site then excitation can exchange with neighboring pigments on the timescale of population loss and no local depletion occurs. We will refer to this scenario as the `weak quenching'€™ regime. Thus a homogeneous distribution of weak quenchers across the membrane will cause a decrease in the excitation diffusion length but will not distort the population transport mechanism from the 2D random walk observed for dark-acclimated leaves. 

To determine whether qE occurs in the strong or weak queching regime, we turned to recent single molecule fluorescence lifetime data on quenched PSII supercomplexes by Gruber, \it et al. \rm \cite{Gruber2016}. Using our structure-based model of energy transfer in PSII supercomplexes \cite{Bennett2013}, we could estimate the timescale of quenching at a thermally populated domain that is consistent with the 50 ps timescale extracted by Gruber et al. \cite{Gruber2016} (Supplementary text). Assuming the quenching site is the lowest energy Chl610-612 domain of LHCII (LHCII-610) \cite{Ruban2007,Kruger2012}, we estimate the underlying timescale of quenching ($\tau_{\textrm{qE}}$) to be 20 ps (Supplementary text, Table S1; Fig.~\ref{fig2}C, $\tau_{\textrm{qE}} \approx 20$ ps). This timescale is in good agreement with semi-empirical calculations of quenching by transport from the LHCII-610 domain to the adjacent lutein ($\tau_{\textrm{qE}} \approx 30$ ps) \cite{Duffy2012}. The 20-30 ps timescale of quenching at the LHCII-610 domain is slow compared to the $\sim$3 ps timescale on which excitation transport exits that domain (Supplementary text). As a result, only a small fraction ($\sim$10\%) of excitation is dissipated during a single visit to an active quenching site, which indicates that LHCII-610 is a weak quencher. The same conclusion is reached when assuming either of the other proposed qE quenching sites (Supplementary text, Table S1): the Chl610-612 domain of the minor light harvesting complexes (mLHC-610) \cite{Ahn2008,Avenson2009} or the Chl601,608,609 domain of LHCII (LHCII-608) \cite{Schlau-Cohen2015}. The fluorescence lifetimes measured on single quenched LHCII complexes are also consistent with a weak quenching regime \cite{Schlau-Cohen2015}. Taken together, current single molecule evidence supports a weak quenching model of qE for all proposed quenching sites.

Our simulations confirm that excitation transport remains a 2D diffusive process in presence of weak qE quenchers and predict that the excitation diffusion length in the PSII membrane decreases from 50 nm to 25 nm during bright light acclimation of an \emph{A. thaliana} leaf. We simplified the multiple states observed in single-molecule measurements \cite{Kruger2012,Schlau-Cohen2015} to a two-state switch model for light harvesting proteins. A quenching site has a probability ($P_{\textrm{qE}}$) of being bound in a conformation in which there is an additional decay pathway with a timescale of $\tau_{\textrm{qE}}$ (Fig.~\ref{fig2}C). We scanned $P_{\textrm{qE}}$ between 0 and 1 and $\tau_{\textrm{qE}}$ between 10 ps and 100 ps to explore any dependence on the specific rate and density of quenchers. Consistent with the weak quenching picture, the fluorescence lifetime measurements  \cite{Sylak-Glassman2016} on light-adapted wild-type \emph{A. thaliana} leaves (blue line, Fig.~\ref{fig2}D) can be equivalently described by a wide range of ($\tau_{\textrm{qE}}$, $P_{\textrm{qE}}$) values with the LHCII-610 site (blue dots, Fig.~\ref{fig2}D and blue area, Fig.~\ref{fig2}E). These equivalent decay curves all correspond to ($\tau_{\textrm{qE}}$, $P_{\textrm{qE}}$) combinations that give rise to an excitation diffusion length ($L_{\textrm{D}}$) of 25 nm (Fig.~S2C). The other proposed sites of quenching are equally capable of reproducing the fluorescence lifetime measurements but with different ($\tau_{\textrm{qE}}$, $P_{\textrm{qE}}$) values (Fig.~S2A). The resulting excitation diffusion length ($L_{\textrm{D}}$), however, remains 25 nm (Fig.~S2B). The decreased $L_{\textrm{D}}$ in the presence of qE completely characterizes, when RCs are closed, the fraction of excitation emitted from chlorophylls as fluorescence, lost to inter-system crossing, and quenched by the activated quenching sites (Fig.~S2D). Thus, the influence of a uniform distribution of weak qE quenchers can be encapsulated by the decrease in the excitation diffusion length (blue area, Fig.~\ref{fig2}B).

\subsection{Influence of a variable excitation diffusion length on photochemical yield.}
We have established that qE modifies the excitation diffusion length in the PSII membrane and that this control parameter is invariant to the precise site of quenching. How does the change in the excitation diffusion length ($L_{\textrm{D}}$) influence the trapping of excitation at open reaction centers? We note that unlike qE quenching sites, RCs are strong quenchers and the net process of excitation transport in the presence of open RCs is described by the combination of $L_{\textrm{D}}$ and the fraction of open RCs ($f_{\textrm{RC}}$). 

Changes in $L_{\textrm{D}}$ affect both the extent to which reaction centers compete with each other for excitations and the fraction of excitations that are available to reaction centers. Competition between reaction centers is the origin of the hyperbolic dependence of the photochemical yield ($\Phi_{\textrm{PC}}$, i.e. the fraction of excitations reaction centers convert to chemical energy by charge separation) on the fraction of open reaction centers ($f_{\textrm{RC}}$) originally measured by Joliot and Joliot \cite{Joliot1964}. The decrease in the fraction of excitations available to reaction centers arises from quenching by active qE sites that are spatially separated from open reaction centers. As excitation diffuses through the membrane, it can be quenched prior to reaching the first reaction center or between visits to reaction centers. The average antenna size of a reaction center ($\sigma_{\textrm{PC}}$, i.e. the photochemical cross-section, Supplementary text), encapsulates both the effects of competition between reaction centers and the reduction in available excitations due to qE  \begin{equation} \label{eq:sigma}
\sigma_{\textrm{PC}}(L_{\textrm{D}}, f_{\textrm{RC}}) = \frac{\Phi_{\textrm{PC}}(L_{\textrm{D}}, f_{\textrm{RC}})\cdot \textrm{N}_{\textrm{ChlA}}}{\textrm{N}_{\textrm{RC}}\cdot f_{\textrm{RC}}},
\end{equation}
where $\textrm{N}_{\textrm{RC}}$ and $\textrm{N}_{\textrm{ChlA}}$ are the number of RCs and chlorophyll A in the membrane, respectively, $\Phi_{\textrm{PC}}$ is the photochemical yield, and $f_{\textrm{RC}}$ is the fraction of open RCs. We measure the antenna size by the number of chlorophyll A equivalents from which the average RC will harvest energy. When $L_{\textrm{D}}$ is very large, we expect the average antenna size of reaction centers in PSII ($\sigma_{\textrm{PC}}$) to be large but constrained by the competition between open reaction centers. 

\begin{figure*}[t!]
\begin{center}
\includegraphics{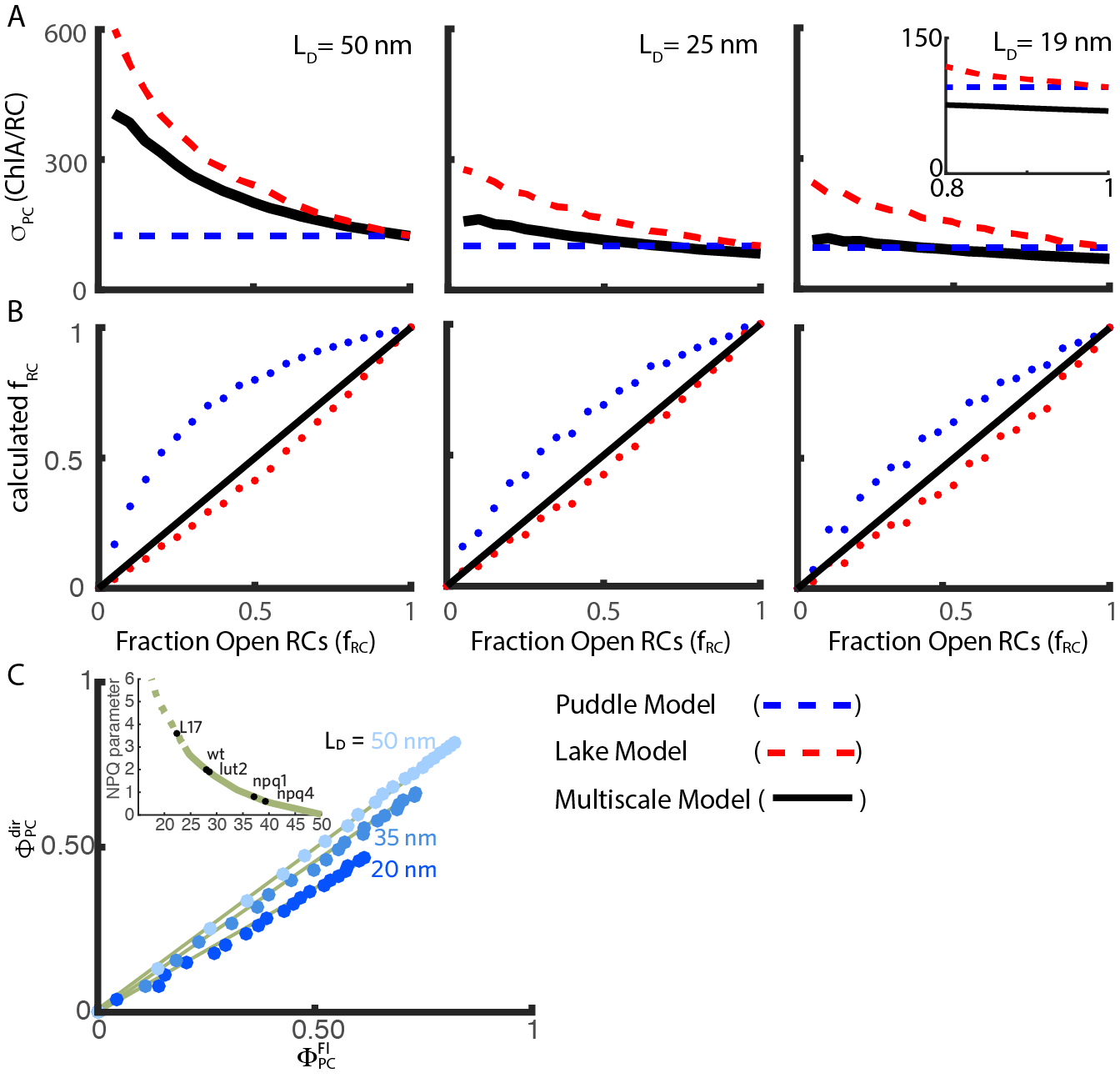}
\caption{\textbf{Lake and puddle models do not capture a variable excitation diffusion length.} We used the chlorophyll fluorescence simulated with the multiscale model of PSII to calculate photochemical cross-section ($\sigma_{\textrm{PC}}$), calculated fraction of open reaction centers ($f_{\textrm{RC}}$), and photochemical yield $\Phi_{\textrm{PC}}^{\textrm{Fl}}$ for the lake and puddle models (Supplementary text). \textbf{(A)} Photochemical cross-section (or average antenna size of a reaction center), $\sigma_{\textrm{PC}}$, as a function of $f_{\textrm{RC}}$, the fraction of open reaction centers used in simulations. The inset on the right panel zooms in on $0.8<f_{\textrm{RC}}<1$.  \textbf{(B)} The calculated fraction of open RCs (calculated $f_{\textrm{RC}}$) plotted against the fraction of open reaction centers used in simulations ($f_{\textrm{RC}}$). \textbf{(C)} The points indicate photochemical yield calculated directly using the multiscale model ($\Phi_{\textrm{PC}}^{\textrm{dir}}$) and the photochemical yield calculated for the lake and puddle models ($\Phi_{\textrm{PC}}^{\textrm{Fl}}$) at a given excitation diffusion length ($L_\textrm{D}$) as function of varying the fraction of open reaction centers ($f_{\textrm{RC}}$). The lines through the points were calculated using the quantitative relationship between the slopes of all such calculated lines and the $L_\textrm{D}$ (Fig.~S3). The green line (dashed and solid) in the inset shows the one-to-one relationship between the so-called NPQ parameter (eq.~\ref{eq:npq}, Methods), and the excitation diffusion length ($L_\textrm{D}$). The black points indicate the measured steady-state values of the NPQ parameter of several qE mutants at 1200 $\mu$mol photons m$^{-2}$ s$^{-1}$: L17, a PsbS overexpressor \cite{Li2002}, and \emph{npq1}, \emph{lut2},  and \emph{npq4}, which are lacking zeaxanthin, lutein, and PsbS, respectively \cite{Niyogi2001}.  The solid parts of the line indicate the range of excitation diffusion lengths determined by the $P_{\textrm{qE}}$ suggested from recent single-molecule measurements of LHCII \cite{Kruger2012} (Supplementary text) and the $\tau_{\textrm{qE}}$ suggested from single molecule measurements of PSII supercomplexes \cite{Gruber2016} and modeling of quenching at the Chl610-612 domain by lutein \cite{Duffy2012} (Supplementary text). \label{fig3}}
\end{center}
\end{figure*}

When a single reaction center is open, the average antenna size for reaction centers ($\sigma_{\textrm{PC}}$) decreases from $\sim$400 to $\sim$100 as the excitation diffusion length ($L_{\textrm{D}}$) goes from 50 nm to 19 nm. Note that when $L_{\textrm{D}} = 19$ nm, the average RC antenna size is smaller than that expected for a reaction center in an isolated PSII supercomplex in the absence of qE. This points to the critical role of entropy in light harvesting by the PSII membrane: while the isolated PSII supercomplex contains many fewer pigments than a 19 nm radius on the membrane, in the membrane many excitations will wander away from the open reaction center rather than lingering close as they must in an isolated PSII supercomplex. As the number of open reaction centers increases, the average reaction center antenna size ($\sigma_{\textrm{PC}}$) will decrease due to competition (black line, Fig. \ref{fig3}A). Reaction centers compete most when the excitation diffusion length is longest ($L_{\textrm{D}} = 50$ nm), as seen by the steep decrease of the average RC antenna size from $\sim$400 to $\sim$100 ChlA when going from 1 open reaction center to all reaction centers open. When the excitation diffusion length is much shorter ($L_{\textrm{D}} = 19$ nm), reaction centers are much closer to independent and only a small change in the average reaction center antenna size is observed between 1 reaction center open and all reaction centers open. These relationships between photochemical yield and fraction of open reaction centers depend solely on the excitation diffusion length, and does not vary with different sites of quenching (Fig.~S2E, Supplementary text). 

\subsection{Coarse-grained model of PSII light harvesting.}

Having established that qE regulates PSII light harvesting by adjusting the excitation diffusion length ($L_{\textrm{D}}$) we now explore how this mesoscopic behavior influences the accuracy of coarse-grained models that are used for (i) interpreting the measurements of variable chlorophyll fluorescence and (ii) describing PSII light harvesting in longer length scale simulations of photosynthetic metabolism. The correct interpretation of variable chlorophyll fluorescence measurements of leaves is critical to measuring changes in the state of PSII during bright light adaptation \cite{Zaks2012}, which is often used as an indicator for photosynthesis as a whole \cite{Porcar-Castell2014}. Likewise, simulating photosynthetic metabolism as a whole requires an efficient description of how qE regulates PSII light harvesting and the resulting flux to downstream reactions \cite{Zhu2013}. Thus, a quantitative understanding of the fitness benefits of qE \it in vivo \rm requires a coarse-grained model of PSII light harvesting that accurately captures the influence of a variable excitation diffusion length.

The lake and puddle models have been widely used since the 1960s \cite{Robinson1966,Lazar1999} to describe PSII light harvesting in the presence of qE, but they coarse-grain over the spatial aspects of competition. In particular, both models assume that the fraction of excitation available to reaction centers is the same as qE activates (i.e. the excitation diffusion length $L_\textrm{D}$ is independent of qE). The lake and puddle models differ, however, in their assumptions about the form of competition between reaction centers. The lake model assumes that all open reaction centers compete equally for each excitation, corresponding to $L_\textrm{D} \rightarrow \infty$ and an inverse relationship between photochemical cross-section and the fraction of open reaction centers (Supplementary text). On the other hand, in the puddle model $L_{\textrm{D}}$ is assumed to be sufficiently short that no reaction centers compete, and thus photochemical cross-section is independent of the fraction of open reaction centers (Supplementary text).

What are the implications of a changing excitation diffusion length on the interpretation of chlorophyll fluorescence using the lake and puddle models? For the coarse-graining of PSII light harvesting for metabolic models? In the following, we use the chlorophyll fluorescence simulated with the multiscale model of PSII to calculate the average RC antenna size (Fig.~\ref{fig3}A), the fraction of open RCs (Fig.~\ref{fig3}B), and the photochemical yield (Fig.~\ref{fig3}C) determined by the lake and puddle models (Supplementary text). These results assess the extent to which the lake and puddle models can recapitulate the multiscale simulation.

Both the lake and puddle models fail to describe the changes in the competition between reaction centers due to a variable excitation diffusion length. The lake model (Fig.~\ref{fig3}A, red dashed line) shows good agreement with the multiscale model when the excitation diffusion length is long ($L_{\textrm{D}}$ = 50 nm), but overestimates the average antenna size when the diffusion length decreases. On the other hand, the absence of competition between reaction centers in the puddle model (Fig.~\ref{fig3}A, blue dashed line) provides a reasonable description when $L_{\textrm{D}}$ = 19 nm, but not when $L_{\textrm{D}}$ = 50 nm. Furthermore, neither coarse-grained model captures the photochemical cross-section ($\sigma_{\textrm{PC}}$) in the intermediate regime appropriate for qE in wild type {\it A. thaliana} exposed to 1200 $\mu$mol photons m$^{-2}$ s$^{-1}$ for 10 minutes (Fig.~\ref{fig2}D,E). The puddle model performs poorly compared to the lake model for determining the fraction of open reaction centers across the physiological range of excitation diffusion lengths (Fig.~\ref{fig3}B) because the dimeric structure of PSII supercomplexes ensures that reaction centers always experience some competition. 

In keeping with the assumption that all excitations are available to reaction centers, both the lake and puddle models overestimate the photochemical yield ($\Phi_{\textrm{PC}}$) in the presence of qE. The vertical offset of the photochemical cross-section ($\sigma_{\textrm{PC}}$) when all reaction centers are open ($f_{\textrm{RC}} = 1$) and $L_{\textrm{D}} = 19$ nm (Fig.~\ref{fig3}A, right, inset) represents a $\sim$30\% error in the photochemical yield predicted by the lake and puddle models compared to the multiscale model. We quantify the error in the photochemical yield extracted from chlorophyll fluorescence ($\Phi_{\textrm{PC}}^{\textrm{Fl}}$, lake/puddle models result in the same photochemical yields \cite{Kramer2004}) by comparing to the yield directly calculated from the multiscale model ($\Phi_{\textrm{PC}}^{\textrm{dir}}$) (dots, Fig.~\ref{fig3}C). At a given excitation diffusion length, there is a linear relationship between $\Phi_{\textrm{PC}}^{\textrm{Fl}}$ and $\Phi_{\textrm{PC}}^{\textrm{dir}}$ ($\Phi_{\textrm{PC}}^{\textrm{dir}} = m(L_{\textrm{D}}) \cdot \Phi_{\textrm{PC}}^{\textrm{Fl}}$) calculated for a range of fraction open reaction centers. The slope $m(L_{\textrm{D}})$ decreases as the excitation diffusion length ($L_{\textrm{D}}$) decreases because a larger fraction of excitations become inaccessible to any reaction center and thus do not enter into the competition between productive photochemistry and loss pathways. Taken together, neither the lake nor puddle model correctly determines the photochemical yield of PSII ($\Phi_{\textrm{PC}}$) or the average antenna size of a reaction center ($\sigma_{\textrm{PC}}$) from chlorophyll fluorescence data across the physiological range of $L_{\textrm{D}}$s, but the lake model offers a reasonable estimate of the fraction of open reaction centers ($f_{\textrm{RC}}$).  

\begin{figure}[t!]
\begin{center}
\includegraphics{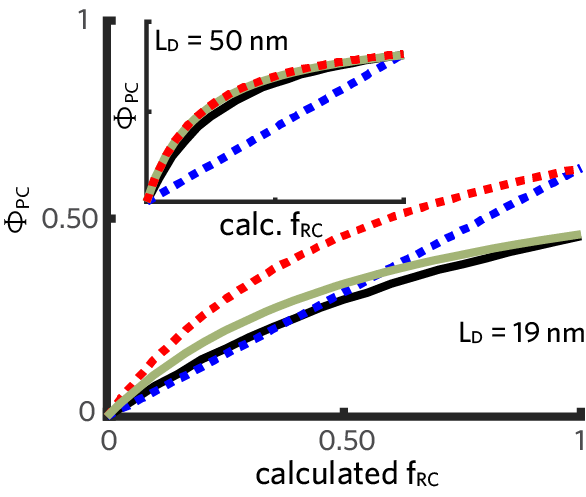}
\caption{\textbf{Correcting the lake model for interpreting chlorophyll fluorescence measurements and larger scale metabolic simulations.} The photochemical yield ($\Phi_{\textrm{PC}}$) simulated using the lake (red dashed line), puddle (blue dashed line), diffusion corrected lake (green solid line) and multiscale model (black line) as a function of the fraction of open reaction centers ($f_{\textrm{RC}}$) calculated when excitation diffusion length ($L_\textrm{D}$) is 19 nm. The inset shows the same plot when $L_\textrm{D} = 50$ nm. \label{fig4}}
\end{center}
\end{figure}

Correcting the lake model to account for an excitation-diffusion-length-dependent fraction of excitations that are inaccessible to reaction centers substantially improves the analysis of chlorophyll fluorescence measurements and the description of PSII light harvesting for longer length scale simulations of photosynthetic metabolism. We fit $m(L_{\textrm{D}})$ (slope of lines in Fig.~\ref{fig3}C) and use it as a lowest-order correction on the photochemical yield calculated using the lake/puddle models ($\Phi_{\textrm{PC}}^{\textrm{Fl}}$) to describe the influence of a diffusion limited pool of excitations available to reaction centers (Fig.~\ref{fig3}C, green lines). To apply this correction in the context of a chlorophyll fluorescence measurement, we first establish an estimate of excitation diffusion length using the one-to-one relationship with the so-called NPQ parameter (Fig.~\ref{fig3}C, inset, Supplementary text) and then apply an $L_{\textrm{D}}$-dependent correction to $\Phi_{\textrm{PC}}^{\textrm{Fl}}$ (Fig.~S3, Supplementary text). This diffusion-corrected lake model offers a substantially improved description of the relationship between the photochemical yield and the fraction of open reaction centers ($f_{\textrm{RC}}$) across a range of excitation diffusion lengths, which is essential for parameterizing higher scale metabolic models (Fig.~\ref{fig4}, green line). We note that the competition between reaction centers causes the hyperbolic dependence of the photochemical yield on the fraction of of open reaction centers and remains overestimated in the diffusion-corrected lake model. Thus, while the diffusion-corrected lake model is a significant improvement, continued effort is needed to develop a coarse-grained model that also accurately captures competition between reaction centers. 

\section*{\large{Discussion}}

There has been considerable debate regarding the site and photophysical mechanism of qE in part because of the potential higher scale consequences of these details. We have demonstrated that the influence of weak quenchers on PSII light harvesting can be described by a reduction in the excitation diffusion length ($L_{\textrm{D}}$) and does not depend on the specific site of quenching. Does qE in fact occur in the weak quenching regime? Single molecule measurements \cite{Gruber2016,Schlau-Cohen2015} and semi-empirical electronic structure calculations \cite{Duffy2012} suggest that the timescale of quenching ($\tau_{\textrm{qE}}$) is slow compared to the timescale of transfer out of the relevant pigment domains, but this is not conclusive. If in fact qE is in the strong quenching regime, the diffusive picture established here would not hold. It is of key importance, then, to determine the photophysical mechanism of quenching and quantitatively establish the timescale of quenching. Further, the emergence of the excitation diffusion length as the central parameter describing PSII light harvesting highlights the need for an experimental determination of this length scale using emerging techniques for spatially resolved ultrafast spectroscopy \cite{Wan2015,Penwell2017}.

Constructing a rigorously coarse-grained model of PSII light harvesting was infeasible in the absence of a structural understanding of the PSII antenna \cite{Lazar1999}. In this absence, the lake and puddle models have succeeded in efficiently capturing the limiting cases of PSII light harvesting in the presence of qE (Fig.~\ref{fig3}A). Here we directly calculate the photochemical cross-section ($\sigma_{\textrm{PC}}$, eq.~\ref{eq:sigma}), using a structure-based excitation transfer model founded on recent developments in x-ray crystallography, ultrafast spectroscopy, and theory of open quantum systems. We demonstrate that neither the lake nor the puddle model correctly captures the influence of a variable excitation diffusion length on PSII light harvesting (Fig.~\ref{fig3}). A simple correction for the excitation-diffusion-length-dependent pool of available excitations is sufficient to substantially improve (i) the calculation of the photochemical yield ($\Phi_{\textrm{PC}}$) from variable chlorophyll fluorescence measurements at both the leaf \cite{Kramer2004} and ecosystem \cite{Porcar-Castell2014} scales (Fig.~\ref{fig3}C) and (ii) the accuracy of the relationship between the fraction of open reaction centers ($f_{\textrm{RC}}$) and the photochemical yield ($\Phi_{\textrm{PC}}$) in the lake model for higher order models of photosynthetic metabolism \cite{Zhu2013} (Fig.~\ref{fig4}). Applying this model to the microsecond rise kinetics of chlorophyll fluorescence could provide insight on the importance of a variable excitation diffusion length and might offer a new perspective of the underlying dynamics of the measurement \cite{Stirbet2011}. 

As discussed above, deciphering the molecular mechanism of qE remains essential. The connection established here from a nanoscale description of excitation dissipation by qE ($\tau_{\textrm{qE}}$, $P_{\textrm{qE}}$) to a mesoscale description of excitation dynamics ($L_{\textrm{D}}$) to an in-field assessment of photoprotective capacity (NPQ parameter) acts as a bridge for comparing measurements made at different length scales and connecting \emph{in vitro} measurements on isolated complexes with \emph{in vivo} measurements on intact leaves. As a proof-of-concept, consider a first estimate of the probability of an activated qE quenching site ($P_{\textrm{qE}}$) of 0.3 from single-molecule measurements of LHCII in conditions mimicking qE (Supplementary text) \cite{Kruger2012} and a timescale of quenching ($\tau_{\textrm{qE}}$) of $\sim$20-30 ps from both single-molecule measurements of PSII supercomplexes \cite{Gruber2016} and semi-empirical electronic structure calculations for the LHCII-610 site \cite{Duffy2012}. The solid line in Fig.~\ref{fig3}, inset, indicates the range of excitation diffusion lengths consistent with these estimates, while the dashed line corresponds to excitation diffusion lengths that are too short. The level of quenching observed in a mutant in which PsbS is overexpressed (L17, Fig.~\ref{fig3} inset) exceeds the \it in vitro \rm bound, which we note is consistent with the recent proposal that LHCII-610 and mLHC-610 quenching sites act in tandem subject to separate regulatory control \cite{DallOsto2017}. 

\section*{\large{Concluding Remarks}}

We have reduced the complexity of PSII light harvesting in the presence of weak quenching, a very heterogeneous process composed of thousands of rate constants, down to a single conceptually rich parameter, the excitation diffusion length in the antenna ($L_\textrm{D}$). We find that in response to fluctuating light intensity qE acts as a `tap' that adjusts the flux of excitation to open reaction centers ($\sigma_{\textrm{PC}}$) via the $L_{\textrm{D}}$. This approach rigorously links the nanoscopic dynamics occurring within pigment-protein complexes to the mesoscopic dynamics of PSII light harvesting that influence CO$_2$ fixation yields in intact leaves. Looking forward, we expect this framework to offer a fertile avenue for resolving the qE mechanism and parameterizing higher scale models of plant function. 

\section*{\large{Acknowledgements}}
The authors are extremely grateful to Aliz{\'e}e Malno{\"e} and Emily Jane Sylak-Glassman for sharing \emph{A. thaliana} PAM data and fluorescence lifetime snapshot data, respectively. DIGB and KA would like to thank Drew Ringsmuth, Andrian Gutu, Aliz{\'e}e Malno{\"e}, Masa Iwai, and Lena Simine for helpful comments on the manuscript. DIGB was supported by CIFAR, the Canadian Institute for Advanced Research, through the Bio-Inspired Solar Energy program, and the Center for Excitonics, an Energy Frontier Research Center funded by the U.S. Department of Energy, Office of Science and Office of Basic Energy Sciences, under Award Number DE-SC0001088. KA was supported by the Howard Hughes Medical Institute through EK O'Shea. KA and GRF were supported by the Director, Office of Science, Office of Basic Energy Sciences, of the U.S. Department of Energy under Contract DE-AC02-05CH11231 and the Division of Chemical Sciences, Geosciences and Biosciences Division, Office of Basic Energy Sciences through Grant DEAC03-76SF000098 (at Lawrence Berkeley National Labs and U.C. Berkeley). This research used resources of the National Energy Research Scientific Computing Center, a DOE Office of Science User Facility supported by the Office of Science of the U.S. Department of Energy under Contract No. DE-AC02-05CH11231. This research used computational time on the Odyssey cluster, supported by the FAS Division of Science, Research Computing Group at Harvard University.
\clearpage

\beginsupplement

\section*{\large{Methods}}
\addcontentsline{toc}{section}{\protect\numberline{}Methods}%

\subsection*{Overview of multiscale model}
\addcontentsline{toc}{subsection}{\protect\numberline{}Overview of multiscale model}%
A model for PSII light harvesting that includes qE quenching requires several components: i) a membrane structure with pigment resolution, ii) a model for energy transfer, iii) a model for electron transfer in the reaction centers, and iv) a model for quenching. Combining these components results in a rate matrix that contains all the rates of transport and loss in the membrane \cite{Zaks2013}. We described i)-iii) in detail in previous work \cite{Bennett2013,Amarnath2016}, and here provide a brief description.

We generated membranes with different organizations of PSII supercomplexes and LHCIIs using Monte Carlo simulations of a 200 nm x 200 nm area containing coarse-grained PSII supercomplex (pill) and LHCII particles (circle) \cite{Schneider2013}. We superimposed the crystal structures of the chlorophyll pigments of the C$_2$S$_2$ PSII supercomplex \cite{Caffarri2009} and LHCII \cite{Liu2004} to establish the locations of all pigments in the membrane. We assumed random orientations of the LHCIIs. The Monte Carlo simulations allow for the energetic attachment of LHCIIs to the PSII supercomplexes in the location where the so-called `medium-bound' LHCIIs can bind, which results in the largest stable form of the PSII supercomplex in plants, the C$_2$S$_2$M$_2$ supercomplex \cite{Kouril2012}. We have not included CP24, and the minor complexes are modeled as LHCII monomers, since the crystal structures of CP24 and CP26 are still lacking. These approximations, which have been justified earlier \cite{Bennett2013}, have led to structure-based predictions of energy transfer that are consistent with data from isolated PSII supercomplexes \cite{Bennett2013} and an intact thylakoid membrane \cite{Amarnath2016}. 

Excitation energy transport rates between domains of strongly coupled chlorophyll ($\sim$3-4 pigments in size) were calculated using Generalized F{\"o}rster theory assuming that excitations thermalize within domains prior to hopping between domains. A quantum-mechanically exact calculation of energy transfer for a system of the size of the thylakoid membrane ($\sim$50,000 pigments) is currently infeasible. Computationally more tractable perturbative methods include modified Redfield theory, in which interactions between electronic states and the vibrational modes of the system are perturbative, and Generalized F{\"o}rster theory, in which the electronic coupling between pigments is perturbative \cite{Ishizaki2010}. The pigment-protein complexes that compose PSII do not lie on either of these extremes, so Renger \cite{Raszewski2008} and Novoderezhkin \cite{Novoderezhkin2011a} pioneered an approach in which pigments are grouped into domains within which transport can be described by Modified Redfield theory, and between which transport is described by Generalized F{\"o}rster theory \cite{Sumi1999,Scholes2000}. We defined domains to maximize the separation of timescales between intra- and inter-domain transfer \cite{Bennett2013}. We can assume instantaneous equilibration within a domain without loss of accuracy in our simulations; however, the domain picture is the most coarse-grained picture that reproduces the dynamics of transport \cite{Bennett2013}. The domains for a monomer of the LHCII trimer are color-coded in the crystal structure in the main text (Fig.~1C, Energy Transfer). The rate of transport between two domains, a donor $d$ and an acceptor $a$, is the Boltzmann-weighted sum of the rates of transport between each pair of excitons in the two domains ($\vert M \rangle$ and $\vert N \rangle$, respectively):
\begin{equation}
\label{eq:gf}
k^{\rm{dom}}_{a \leftarrow d} = \sum_{\substack{ \vert M \rangle \in d \\ \vert N \rangle \in a}} k_{ N \leftarrow M}P^{(d)}_{M},
\end{equation}
where $P^{(d)}_{M} = \frac{e^{-E_M/k_BT}}{\sum_{\substack{\vert M \rangle \in d} \\ }e^{-E_M/k_BT}} $. The rate between two excitons was calculated using Generalized F{\"o}rster theory,
\begin{equation}
\label{eq:overlap}
k_{N \leftarrow M} =  \frac{\vert V_{M, N} \vert^2}{\hbar^2 \pi}  \int_0^{\infty} \textrm{d}t A_{N}(t) F^{\ast}_{M}(t),
 \end{equation}
 where $\vert V_{M, N} \vert^2$ is the electronic coupling between the two excitons and $\int_0^{\infty} \textrm{d}t A_{M}(t) F^{\ast}_{N}(t)$ is the overlap integral between the absorption of the acceptor and the fluorescence of the donor. This model of excitation transport shows good agreement with more exact methods for simulating the excitation population dynamics \cite{Roden2016,Kreisbeck2014}. Here we inhomogeneously average the rates of transport before calculating the dynamics, which offers substantial computational savings. This averaging does not substantially affect excitation dynamics \cite{Amarnath2016}. In this work we additionally lowered any rate constants greater than 2 ps$^{-1}$, which arise due to imperfect overlay of the pigments on the Monte Carlo simulations \cite{Amarnath2016}, to 2 ps$^{-1}$. This tempering does not influence the simulated photochemical yield or fluorescence decay \cite{Amarnath2016}.

We modeled electron transfer using a simple phenomenological kinetic model with three rate constants that describe both reversible and irreversible charge separated states:
\begin{equation}
\vcenter {\hbox{\includegraphics[scale=0.8]{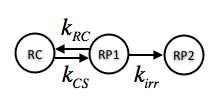}}}.
\end{equation}
Here, RC is the reaction center domain composed of the 6 pigments of the reaction center. The ``radical pair'' states RP1 and RP2 and the rate constants  $k_{RC}$, $k_{CS}$, and $k_{irr}$ are used to model the electron transfer steps in the reaction center. RP1 and RP2 are non-emissive states that do not necessarily have a direct physical analog with charge-separated states in the reaction center. Rather, this approach allows us to describe the overall process of a reversible charge separation step followed by an irreversible step and establish the approximate timescales of these events relative to energy transfer in the light harvesting antenna. The rates were previously parameterized using fluorescence decays of PSII supercomplexes of different sizes \cite{Caffarri2011}, with $\tau_{CS} = 0.64$ ps, $\tau_{RC} = 160$ ps, and $\tau_{irr} = 520$ ps ($k=1/\tau$) \cite{Bennett2013}.  These same three parameters can fit data from PSII supercomplexes with a range of antenna sizes \cite{Bennett2013} and the thylakoid membrane\cite{Amarnath2016} without any alteration or refitting (Fig.~2A, red lines, main text). Closed RCs cannot perform irreversible charge separation,  so $k_{irr} \rightarrow 0 $ and only two kinetic rates are incorporated to describe a reversible charge separated state. We determined $\tau_{CS} = 310$ ps and $\tau_{RC} = 458$ ps by fitting to fluorescence decay data from leaves with closed RCs \cite{Amarnath2016} (Fig.~2A, black lines, main text).

We modeled qE by adding a 1st-order rate of quenching at a qE site (domain) in activated antenna complexes. Each protein housing a qE site can occupy two conformational states, one inactive and one active, and it occupies the active state with probability $P_{\textrm{qE}}$. This simple two-state switch is a simplification of the multiple states observed in single-molecule data \cite{Kruger2012,Schlau-Cohen2015}. In the active state, there is a time scale of excitation dissipation, $\tau_{\textrm{qE}} = 1/k_{\textrm{qE}} $, from the qE site. This effectively coarse-grains the photophysical mechanism of quenching, of which several have been postulated \cite{Ruban2007,Miloslavina2008,Holt2005}, into a single 1st order rate constant. In the main text we primarily consider the Chl610-612 domain on LHCII as it has proposed by multiple groups to be a site of quenching \cite{Ruban2012a}. We only allow unbound LHCII (circles in Fig.~1B of main text) to quench excitation. However, the overall effect of weak quenching - that excitation visits on average several active qE sites before being quenched (see below) - means that this assumption does not effect any of the main results. Moreover, we show in Fig.~S2 that moving the site of quenching to the proposed Chl601,608,609 domain in LHCII (LHCII-608) \cite{Schlau-Cohen2015} or the Chl610-612 domain in the minor light harvesting complexes (mLHC-610) \cite{Ahn2008,Avenson2009} does not change the behavior of quenching described in the main text. Further, we have considered the role of membrane organization. There have been reports of the segregation of PSII supercomplexes and LHCII during qE \cite{Betterle2009,Johnson2011}, though this is not established. Fitting fluorescence lifetime data with quenching on LHCII-610 in segregated membranes results in a similar area of best-fit as for mixed membranes, but shifted to the right (Fig.~S1). Despite this similarity, the spatial heterogeneity of quenchers means that the excitation diffusion length by itself is insufficient as a simplifying concept, and we do not pursue the coarse-graining of quenching in segregated membranes here.

The overall picture of PSII light harvesting from our model is as follows. The nodes of the energy transfer network are domains of $\sim$3-4 tightly coupled chlorophylls. At every domain, excitation can be transferred to nearby domains or dissipated by fluorescence ($\tau_{\textrm{Fl}} = 16$ ns) or non-radiative decay ($\tau_{\textrm{nr}} = 2$ ns). Active qE quenching domains have an additional pathway of dissipation with a timescale $\tau_{\textrm{qE}}$. The reaction center domains can additionally dissipate excitation by the final irreversible step from the RP1 kinetic compartment to the RP2 compartment.

\subsection*{Running simulations}
\addcontentsline{toc}{subsection}{\protect\numberline{}Running simulations}%
The result of the model building in the previous section is a $\sim$11000 x 11000 rate matrix ($K$) that describes the total kinetic network of PSII light harvesting. We use two simulation approaches. We use a finite difference calculation of population dynamics to simulate fluorescence lifetime curves. We use Kinetic Monte Carlo to calculate yields for the different decay pathways and the excitation diffusion length scale. For simulations that involve the presence of qE sites or fractionally open RCs, we average between 10 and 50 different configurations of qE quenchers and open RCs on the membrane, where in each simulation the probability of a given quencher being active is $P_{\textrm{qE}}$ and the probability of a given RC being open is $f_\textrm{RC}$. In theory, the influence of closing one RC modifies the probability that an adjacent RC is closed. We have simulated this previously but found the effect to be small enough that it is neglected here. 

\subsubsection*{Simulating Fluorescence Lifetime Curves}
In a master equation formalism, the time-dependence of excitation population ($P(t)$) is determined entirely by $K$: 
\begin{equation}
\label{eq:master}
\dot{P}(t) = KP(t).
\end{equation}
We calculated $P(t)$ using the finite difference method with a time-step of 10 fs which was found to be well converged. $P(0)$ in all cases was proportional to the number of ChlA in the domain.

\subsubsection*{Monte Carlo Simulations}
On a given membrane, a kinetic Monte Carlo scheme was also used to determine the yields of the different decay pathways using 5,000 trajectories with an initial domain sampled with a probability proportional to the number of ChlA in the domain. Kinetic Monte Carlo was also used to calculate the diffusion length scale. In this case, the simulation is done in two steps. First the one-over-e time - the time at which the fraction of the unquenched excitation becomes equal to $e^{-1}$ - is calculated. Then the average absolute displacement from the initial domain at the one-over-e time is calculated for 5,000 trajectories. For $L_{\textrm{D}}$ calculations, the ChlA initial condition is modified to only allow excitation starting in 40 nm x 40 nm patch at the center of the membrane to ensure minimal effects from the boundaries of the membrane. 

\subsection*{Calculation of chlorophyll fluorescence parameters from simulations}
\addcontentsline{toc}{subsection}{\protect\numberline{}Calculation of chlorophyll fluorescence parameters}%
The primary functional outputs of PSII - the yield of photochemistry in the reaction centers, the extent of qE, and the fraction of open RCs are typically calculated from variable chlorophyll fluorescence yield ($\Phi_{\textrm{Fl}}$, see Fig.~1A, main text) measurements using the lake and puddle models \cite{Baker2008,Kramer2004,Ahn2009}. The relative changes in $\Phi_{\textrm{Fl}}$ are typically measured using pulsed amplitude modulated (PAM) fluorescence. PAM fluorescence consists of three light sources: a dim light for measuring chlorophyll fluorescence without perturbing PSII function, an actinic light that mimics changes in natural sunlight intensity and can (de)activate qE, and brief ($<$1 s) high light pulses that close (saturate) all the reaction centers. Prediction of various PSII outputs in the context of the lake or puddle model follows from arranging chlorophyll fluorescence responses to the actinic light and the saturating flashes into equations. For a full discussion of PAM fluorescence and how it is used to monitor PSII function, please see refs.~\cite{Baker2008,Kramer2004}. Here, we simply list the main chlorophyll fluorescence parameters with the state of PSII they correspond to.
\begin{itemize}
  \item Fm: $\Phi_{\textrm{Fl}}$ with no qE activated and all RCs closed
  \item Fm': $\Phi_{\textrm{Fl}}$ with qE activated and all RCs closed
  \item Fs: $\Phi_{\textrm{Fl}}$ with qE activated and some fraction of reaction centers open 
  \item Fo': $\Phi_{\textrm{Fl}}$ with qE activated and all reaction centers open
\end{itemize}

Using our bottom-up multiscale model we can directly calculate the photochemical yield ($\Phi_{\textrm{PC}}^{\textrm{dir}}$) as a function of arbitrary sites of qE, different ($\tau_{\textrm{qE}}$, $P_{\textrm{qE}}$) combinations, and different fractions of open RCs ($f_{\textrm{RC}}$). To compare simulations using our model to the predictions of the lake and puddle models, we calculated the chlorophyll fluorescence parameters from PAM fluorescence listed above from the corresponding fluorescence yields calculated with our simulations. $\Phi_{\textrm{Fl}}$ depends on the levels of qE and the fraction of open RCs. In the main text, we show that the level of qE is described by the excitation diffusion length ($L_{\textrm{D}}$) and the fraction of open RCs, $f_{\textrm{RC}}$. The $L_{\textrm{D}}$ is 50 nm with no qE active, and less than 50 nm if any qE is active.

We calculated the fraction of open RCs predicted by the lake ($q_{\textrm{L}}$, red dots in Fig.~3B, main text) and puddle ($q_{\textrm{P}}$, blue dots in Fig.~3B, main text) models as follows:
\begin{equation}
\label{eq:qP}
q_{\textrm{P}} = \frac{\textrm{Fm'} - \textrm{Fs}}{\textrm{Fm'} - \textrm{Fo'}} = \frac{\Phi_{\textrm{Fl}}(L_{\textrm{D}}, f_{\textrm{RC}} = 0) - \Phi_{\textrm{Fl}}(L_{\textrm{D}}, f_{\textrm{RC}})}{\Phi_{\textrm{Fl}}(L_{\textrm{D}}, f_{\textrm{RC}} = 0) - \Phi_{\textrm{Fl}}(L_{\textrm{D}}, f_{\textrm{RC}} = 1)} 
\end{equation}

\begin{equation}
\label{eq:qL}
q_{\textrm{L}} = q_{\textrm{P}}\frac{\textrm{Fo'}}{\textrm{Fs}} = q_{\textrm{P}}\frac{\Phi_{\textrm{Fl}}(L_{\textrm{D}}, f_{\textrm{RC}}=1)}{\Phi_{\textrm{Fl}}(L_{\textrm{D}}, f_{\textrm{RC}})} 
\end{equation}

We calculated photochemical yield as predicted by the lake and puddle models (denoted $\Phi_{\textrm{PC}}^{\textrm{Fl}}$ in the main text, and typically denoted as  $\Phi_{\textrm{II}}$ \cite{Kramer2004}) as follows:
\begin{equation}
\begin{aligned}
\label{eq:phiII}
\Phi_{\textrm{PC}}^{\textrm{Fl}}(L_{\textrm{D}},f_{\textrm{RC}}) &= \frac{\Phi_{\textrm{Fl}}(L_{\textrm{D}}, f_{\textrm{RC}} = 0) - \Phi_{\textrm{Fl}}(L_{\textrm{D}}, f_{\textrm{RC}})}{\Phi_{\textrm{Fl}}(L_{\textrm{D}}, f_{\textrm{RC}} = 0)} \\
\Phi_{\textrm{II}} &= \frac{\textrm{Fm'} - \textrm{Fs}}{\textrm{Fm'}}
\end{aligned}
\end{equation}


Lastly, we calculated the NPQ parameter, which is used to measured the extent of activation of any NPQ mechanisms including qE, as follows: 

\begin{multline}
\label{eq:npq}
\textrm{NPQ}(L_{\textrm{D}}) = \\ \frac{\Phi_{\textrm{Fl}}(L_{\textrm{D}}= 50 \textrm{~nm}, f_{\textrm{RC}} = 0) - \Phi_{\textrm{Fl}}(L_{\textrm{D}},f_{\textrm{RC}} = 0)}{\Phi_{\textrm{Fl}}(L_{\textrm{D}}, f_{\textrm{RC}} = 0)} \\ = \frac{\textrm{Fm} - \textrm{Fm'}}{\textrm{Fm'}}
\end{multline}



\subsection*{Fluorescence lifetime snapshot data analysis}
\addcontentsline{toc}{subsection}{\protect\numberline{}Fluorescence lifetime data analysis}%
Fluorescence lifetime snapshots were measured during the application of a dark-light-dark actinic light sequence on dark-adapted wild-type leaves of \emph{A. thaliana} \cite{Sylak-Glassman2016}. The light period was 10 min, with an actinic light intensity of 1200 $\mu$mol photons m$^{-2}$ s$^{-1}$. The snapshots were taken under light conditions in which the reaction centers were closed. Each decay was fit to a sum of three exponential decays, such that the sum of the amplitudes of each decay was normalized to 1. The shortest decay had a time constant that varied between 65 and 85 ps, the middle decay had a time constant that varied between 480 and 1020 ps, and the longest decay had a time constant that varied between 1.2 and 2.3 ns.  The amplitude of the shortest component before actinic light exposure was taken to be the photosystem I (PSI) contribution to all decays. At this time point we assumed no NPQ had turned on and that the only contribution to the shortest lifetime component was from PSI \cite{Ihalainen2005}.  This PSI amplitude was subtracted from the amplitude of shortest decay component at all other time points, since we assume that the PSI contribution to the lifetime does not change during the actinic light sequence. The amplitudes at each time point were then renormalized to sum to 1, resulting in the PSII component of the fluorescence decay. Fig.~2A of the main text indicates the average lifetime PSII component of the fluorescence decay over the course of the the dark-light-dark sequence. Fig.~2D of the main text (black line) indicates the PSII component of the fluorescence decay taken after 6.5 min of light exposure.

\begin{figure*}
\begin{center}
\includegraphics{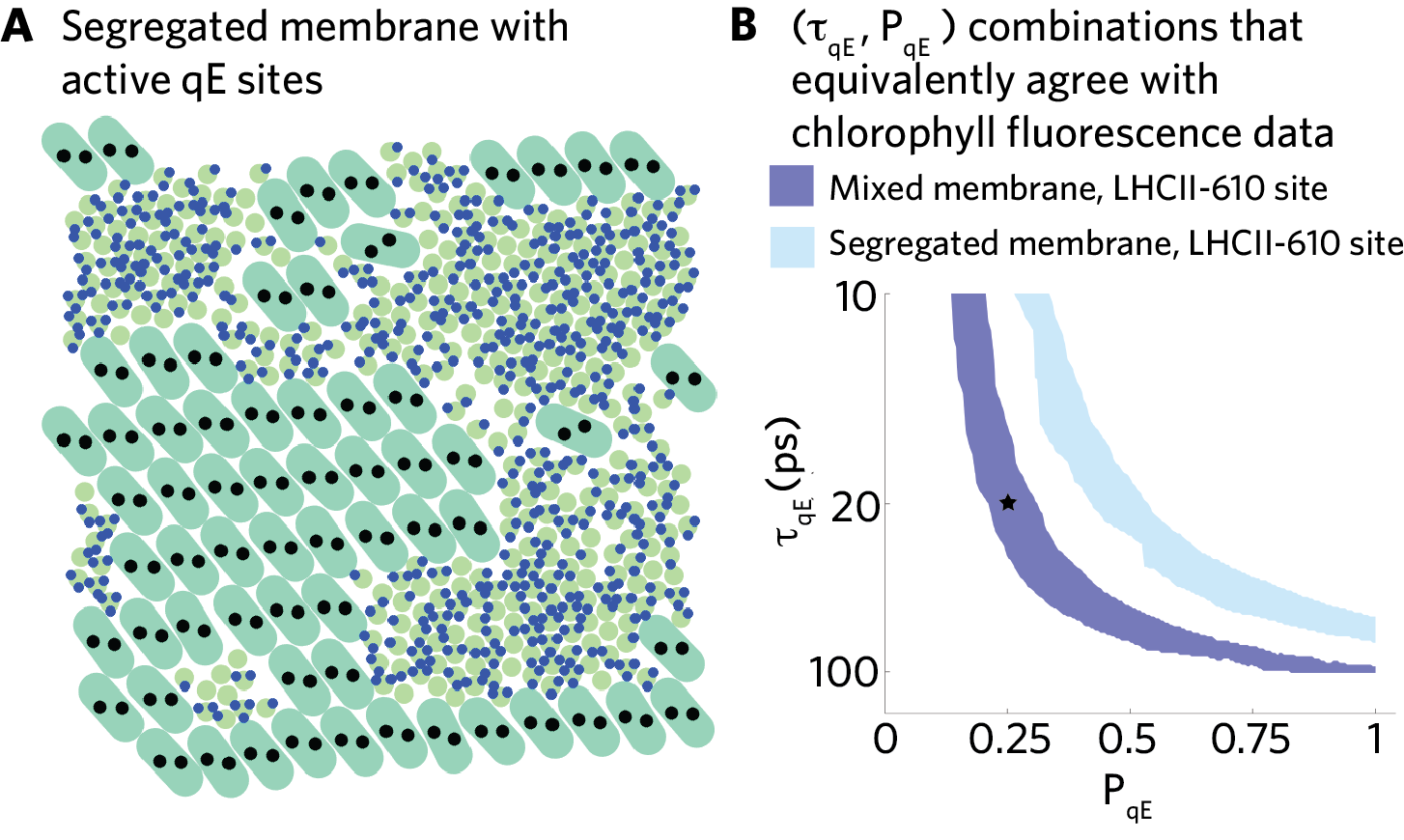}
  \caption{\textbf{Multiple ($\tau_{\textrm{qE}}$, $P_{\textrm{qE}}$) combinations that are consistent with fluorescence lifetime data for a spatially heterogeneous distributions of active qE sites.} \textbf{(A)} Example of a membrane in which the LHCIIs and the PSII supercomplexes are spatially segregated, with all reaction centers closed (black filled circles) \cite{Amarnath2016}. The quenching site is 610-612 domain of LHCII, and activated sites are shown as filled blue dots. \textbf{(B)} The ($\tau_{\textrm{qE}}$, $P_{\textrm{qE}}$) combinations that are consistent with the fluorescence lifetime data (see Methods, Fluorescence lifetime snapshot data analysis for how this was determined) shown in Main Text, Fig.~2D, black line. The fluorescence decay indicated by the solid blue line in Fig.~2D of the main text is for the $\tau_{\textrm{qE}}$, $P_{\textrm{qE}}$ combination shown with a star. Segregation of active quenching sites (light blue) shifts such combinations to the right compared to the mixed case (blue) because of less effective quenching of excitation initiated in the PSII supercomplexes.  \label{fig:seg}}
  \addcontentsline{toc}{subsection}{\protect\numberline{}Fig. S1}%
\end{center}
\end{figure*}

\begin{figure*}
\includegraphics{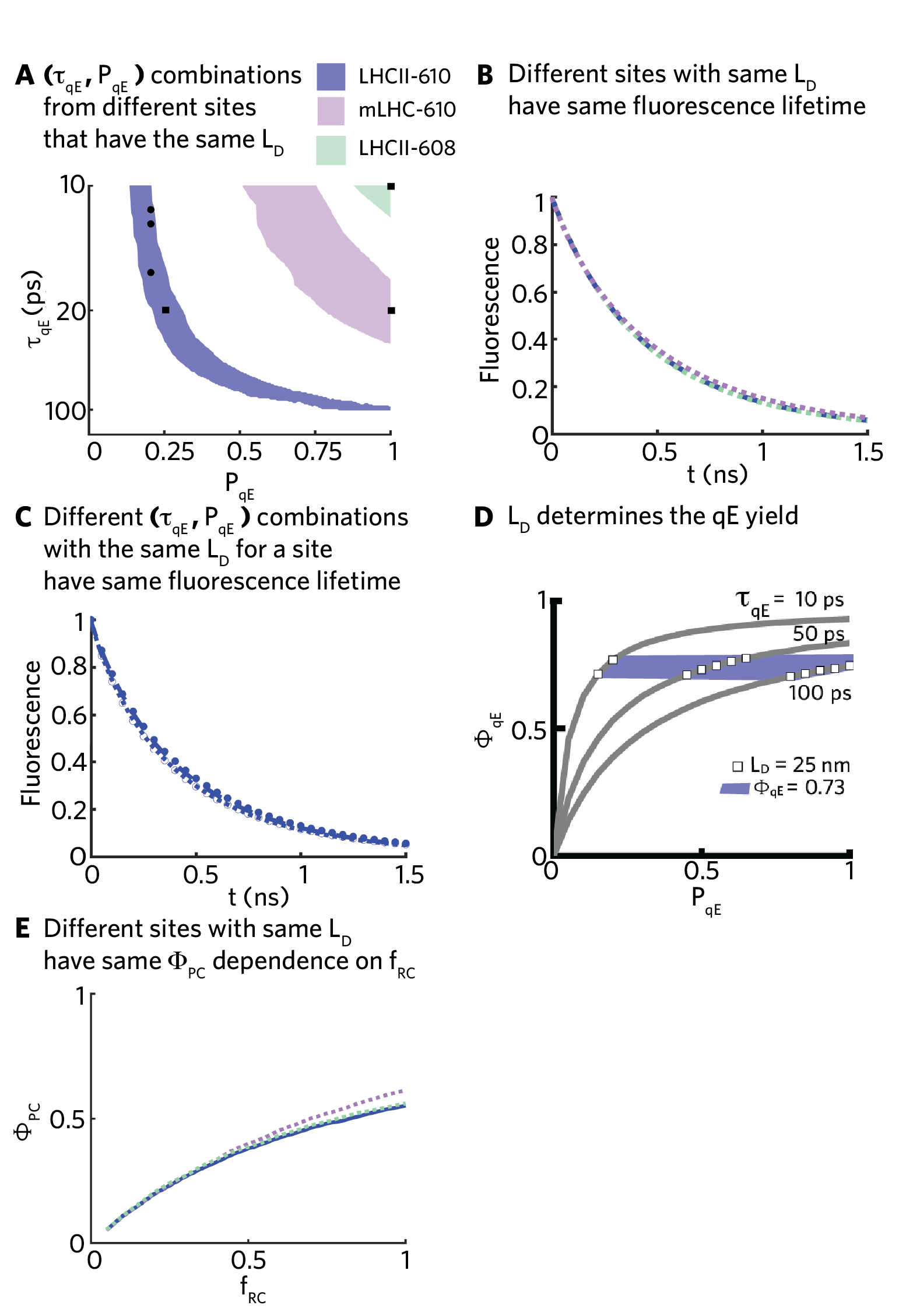}
  \caption{\textbf{The excitation diffusion length ($L_{\textrm{D}}$) determines PSII light harvesting function regardless of the proposed site of quenching.}  \textbf{(A)} The ($\tau_{\textrm{qE}}$, $P_{\textrm{qE}}$) combinations that are consistent with the fluorescence lifetime data (Main Text, Fig.~2D, black line; see Methods, Fluorescence lifetime snapshot data analysis for how this was determined)  also have an $L_{\textrm{D}}$ of $25 \pm 1$ nm independent of the proposed sites of qE. Such combinations for the minor light harvesting complex Chl610-612 (mLHC-610) site (light purple) are shifted to the right relative to those for LHCII-610 (blue) because there are fewer mLHCII-610 sites in the membrane. Only one ($\tau_{\textrm{qE}}$, $P_{\textrm{qE}}$) combination for the LHCII Chl601,608,609 domain (LHCII-608, light blue green) has an $L_{\textrm{D}}$ of $25 \pm 1$ nm because it contains ChlB and has a higher effective free energy than Chl610-612. \textbf{(B)} The fluorescence decays of the points indicated by squares in (A). The colors are consistent with the legend in (A). \textbf{(C)} The fluorescence decays of the points the blue area in (A). \textbf{(D)} Quenching yield ($\Phi_{\textrm{qE}}$) as a function of $P_{\textrm{qE}}$ for $\tau_{\textrm{qE}}$ values ranging from 10 ps to 100 ps. The white squares indicate the calculated points with an $L_\textrm{D}$ of $25\pm 2.5$ nm. The blue region indicates  $\Phi_{\textrm{qE}} = 0.73 \pm 0.05$.  \textbf{(E)} The photochemical yield ($\Phi_{\textrm{PC}}$) as a function of the fraction of open RCs ($f_{\textrm{RC}}$) for the three sites with the qE state indicated by the squares in (A). The LHCII-610 site is indicated by the solid line, while the LHCII-608 and mLHC-610 sites are indicated by dotted lines with all three colors indicated in the legend of (A). \label{fig:isosite}}
  \addcontentsline{toc}{subsection}{\protect\numberline{}Fig. S2}%
\end{figure*}

In Fig.~\ref{fig:seg}B and Fig.~\ref{fig:isosite}A we determine combinations of ($\tau_{\textrm{qE}}$, $P_{\textrm{qE}}$) that have a minimal error with the fluorescence decay indicated by the black line in Fig.~2D of the Main Text. We calculated the error between our simulation ($F_{\mathrm{model}}(t;f_{\textrm{RC}}=0, k_{\textrm{qE}},P_{\textrm{qE}})$) and the data ($F_{\mathrm{data}}(t;T=6.5\textrm{~min})$) using two methods. Combinations deemed to agree with the data had an error below thresholds for both methods. First, we calculated the amplitude weighted error, which is most sensitive to agreement at short times $t$:

\begin{multline}
\label{eq:awerror}
\textrm{Amplitude Weighted Error} = \\ \sum_t F_{\mathrm{data}}(t;T=6.5\textrm{~min}) (F_{\mathrm{model}}(t;f_{\textrm{RC}}=0, \tau_{\textrm{qE}}, P_{\textrm{qE}}) - \\ F_{\mathrm{data}}(t;T=6.5\textrm{~min}))^2.
\end{multline}
Second, we calculated the fluorescence yield ratio error using the following equation:
\begin{multline}
\label{eq:yieldratioerror}
\textrm{Yield Ratio Error} = \\ \textrm{abs}\Biggl( \frac{\sum_t F_{\mathrm{data}}(t, T=6.5 \textrm{~min})}{ \sum_t F_{\mathrm{data}}(t, T= -2 \textrm{~min})} - \\ \frac{ \Phi_{\mathrm{Fl,model}}(f_{\textrm{RC}}=0, \tau_{\textrm{qE}},P_{\textrm{qE}})}{  \Phi_{\mathrm{Fl,model}}(P_{\textrm{open}}=0, k_{\textrm{qE}}=1/\tau_{\textrm{qE}} = 0)} \Biggl).
\end{multline}
The areas indicated in Fig.~\ref{fig:seg}B and Fig.~\ref{fig:isosite}A indicate ($\tau_{\textrm{qE}}$, $P_{\textrm{qE}}$) combinations that predict a fluorescence decay with an amplitude weighted error less than 0.30 and a yield ratio error less than 0.15. $F_{\mathrm{model}}(t)$ and $\Phi_{\mathrm{Fl,model}}$ were calculated as described in ref. \cite{Bennett2013}. We calculated the error for each combination of $P_{\textrm{qE}}=0:0.05:1$ and $k_{\textrm{qE}}= 1/\tau_{\textrm{qE}} = 0:0.005:0.1 \textrm{~ps}^{-1}$ and interpolated intermediate values.

\section*{\large{Supplementary text}}
\addcontentsline{toc}{section}{\protect\numberline{}Supplementary text}%

\subsection*{Weak quenching: evidence from single-molecule measurements}
\addcontentsline{toc}{subsection}{\protect\numberline{}Weak quenching: evidence from single-molecule measurements}%

We estimate the rate of quenching appropriate to each proposed quenching site based on the observation of excitation lifetimes in single-molecule measurements of quenched supercomplexes \cite{Gruber2016}. Using Monte Carlo simulations where excitations instantaneously thermalized within protein compartments (e.g. LHCII monomers or minor light harvesting complexes CP26/CP29), Gruber \it{et al.} \rm isolated an effective quenching rate of 50 ps in quenched antenna proteins \cite{Gruber2016}. Since excitations are assumed to be in equilibrium within a protein compartment, the specific identity of the quenching site (domain) will modify the underlying domain-level quenching rate that matches this result - i.e. a domain that is less populated at equilibrium will require a faster qE quenching rate than its more populated counter part to achieve the same effective quenching rate for the whole protein compartment. Table \ref{tab:quenching} shows the extracted domain quenching rates that reproduce the Monte Carlo protein rates at equilibrium for each of the quenching sites considered here: the lowest-energy domain of LHCII (Chl 610-612, LHCII-610), a ChlB containing domain in LHCII (Chl 601-608-609, LHCII-608) \cite{Schlau-Cohen2015}, and Chl610-612 domain in the the minor light harvesting complexes (mLHC-610) \cite{Ahn2008,Avenson2009}. We note that the rate of quenching extracted from the single-molecule evidence for the LHCII-610 quenching site ($\approx 20$ ps) matches well with semi-empirical calculations of quenching in the LHCII-610 domain by transport to the adjacent lutein ($\tau_{\textrm{qE}} \approx 30$ ps) \cite{Duffy2012}. 

\begin{table}
\begin{center}
    \begin{tabular}{ |c|c|c|c|c|c|}
    \hline
                                & $\tau_{\textrm{qE}}^{\textrm{domain}}$ (ps) & $\tau_{\textrm{dwell}}^{\textrm{median}}$  (ps) & $f_ {\textrm{qE}}^{\textrm{median}} $ & $f_ {\textrm{qE}}^{\textrm{min}} $ & $f_ {\textrm{qE}}^{\textrm{max}} $\\  \hline
             LHCII-610   & 21 & 2.7 & 0.11 & 0.01 & 0.17\\
             LHCII-608   & 1.9 & 0.31 & 0.14 & 0.07 & 0.14\\
             mLHC-610  & 21 & 2.2 & 0.10 & 0.01 & 0.16\\ \hline
    \end{tabular}
\end{center}
\caption{Quenching timescales matched against those extracted from single-molecule measurements ($\tau_{\textrm{qE}}^{\textrm{domain}}$) in ps, the median dwell time for an excitation in the quenching domain ($\tau_{\textrm{dwell}}^{\textrm{median}}$) in ps, and the median, minimum, and maximum fraction of excitations that are dissipated on one visit to an active quenching site ($f_ {\textrm{qE}}^{\textrm{median}} $, $f_ {\textrm{qE}}^{\textrm{min}} $, and $f_ {\textrm{qE}}^{\textrm{max}} $ respectively).}
\label{tab:quenching}
\end{table}

In a diffusive picture, quenching can occur in two limits: weak quenching occurs when the fraction of excitation quenched on each visit to a quenching ($f_{\textrm{qE}}$) is $\ll1$ and strong quenching when $f_{\textrm{qE}} \approx 1$. Weak quenchers modify diffusive transport by decreasing the diffusion length scale of an excitation in the membrsne. We calculate $f_{\textrm{qE}}$ as
\begin{equation}
\label{eq:fqe}
f_{\textrm{qE}} = \frac{\tau_{\textrm{qE}}^{-1}}{\tau_{\textrm{qE}}^{-1} + \tau_{\textrm{dwell}}^{-1}},
\end{equation}
where $\tau_{\textrm{qE}}$ is the timescale of quenching from an active qE site, and $\tau_{\textrm{dwell}}$ is the dwell time of an excitation at the quenching domain. $\tau_{\textrm{dwell}}$ is simply the inverse of the sum of all rates out of the quenching domain.

For all proposed quenching sites the value of $f_{\textrm{qE}}$ using the $\tau_{\textrm{qE}}^{\textrm{domain}}$ extracted from the single-molecule estimates is approximately 0.1 and, therefore, firmly in the weak quenching regime. In the main text and the following, we scan $\tau_{\textrm{qE}}$  between 10 ps and 100 ps to explore any dependence on the specific quenching rate. This range of rates is in keeping with the weak quenching dynamics for all of the proposed quenching sites. A faster rate of quenching could be explored from the LHCII-608 site, though as described in the main text and below, this will not shift the functional behavior which for weak quenchers is determined entirely by the excitation diffusion length. 

\subsection*{Weak quenchers in the presence of open reaction centers}
\addcontentsline{toc}{subsection}{\protect\numberline{}Weak quenchers in the presence of open reaction centers}%
A spatially uniform distribution of weak quenchers in the presence of open reaction centers (RCs), which behave as strong quenchers, can be summarized by the excitation diffusion length through the antenna. For both LHCII-610 and LHCII-608, the spatial distribution of quenchers is uncorrelated with open RCs and homogeneously distributed over the PSII enriched portion of the thylakoid membrane. Fig.~S2E shows that at a fixed excitation diffusion length ($L_{\textrm{D}}$) the photochemical yield as a function of the fraction of open RCs is identical for LHCII-610 (blue solid line) and LHCII-608 (green dashed line). At low fraction of open RCs, mLHC-610 (purple dashed line) also shows identical behavior but deviates slightly with increasing fraction of open RCs. Unlike LHCII-610 and LHCII-608, mLHC-610 arises only in the PSII supercomplexes and therefore co-localizes with RCs. At low fraction of open RCs, this results in only a weak correlation because most mLHC-610 sites will co-localize with closed RCs leading to an effectively uncorrelated spatial distribution of quenchers and open RCs. With the fraction of open reaction centers equal to 1, however, every mLHC-610 site is co-localized with an open RC. Since RCs act as strong quenchers and qE sites are weak quenchers, the resulting competition between quenchers is slightly more favorable for open RCs, leading to the approximate 10\% increase in photochemical yield for mLHC-610 quenching sites. 

\subsection*{Photochemical cross-section}
\addcontentsline{toc}{subsection}{\protect\numberline{}Photochemical cross-section}%
We define the photochemical cross-section ($\sigma_{\textrm{PC}}$, or average antenna size for a given reaction center), as
\begin{equation}
\label{eq:sigmaMS}
\sigma_{\textrm{PC}} = \frac{\Phi_{\textrm{PC}}}{f_{\textrm{RC}}} \frac{\textrm{N}_{\textrm{ChlA}}}{\textrm{N}_{\textrm{RC}}},
\end{equation}
where $\Phi_{\textrm{PC}}$ is the photochemical yield, $f_{\textrm{RC}}$ is the fraction of open RCs, $\textrm{N}_{\textrm{ChlA}}$ is the number of ChlA in the membrane, and $\textrm{N}_{\textrm{RC}}$ is the number of RCs in the membrane.
The antenna increase the effective absorption cross-section of a reaction center by being able to harvest sunlight and transport it efficiently to a reaction center. $\sigma_{\textrm{PC}}$ is the portion of the total absorption cross-section that contributes to productive charge separation at one RC in units of ChlA cross-section. Viewed another way, each ChlA contributes some fraction of excitation to charge separation at a given reaction center. ChlA closer to a reaction center will contribute more, while those farther away will contribute less. $\sigma_{\textrm{PC}}$ is simply the sum over all such contributions for a given reaction center.

It is informative to consider the values of $\sigma_{\textrm{PC}}$ in different limits. In the absence of antenna $\sigma_{\textrm{PC}}$ is simply the number of pigments in the RC domain (i.e. six). When a single RC is open in an intact membrane, $\sigma_{\textrm{PC}}$ decreases from 545 to 115 as $L_\textrm{D}$ decreases from 50 nm to 25 nm. As the fraction of open RCs ($f_{\textrm{RC}}$) increases there is additional competition between open RCs that are within $\sim 1$ $L_\textrm{D}$ (excitation diffusion length) of each other. $\sigma_{PC}$ implicitly incorporates the finite diffusion length and the resulting competition between open RCs and quenching pathways. It these spatial effects of diffusion on competition that could not be taken into account with previous definitions of cross-section developed prior to a structural understanding of the PSII membrane (e.g., \cite{Malkin1981}).

The usefulness of $\sigma_{\textrm{PC}}$ is particularly salient in context of quantitatively differentiating the lake and the puddle models. We use the definitions for $\Phi_{\textrm{PC}}$ described in ref.~\cite{Kramer2004}. In the lake model,
\begin{equation}
\label{eq:sigmaL}
\sigma_{\textrm{PC}} = \left( \frac{k_{\textrm{PC}}}{k_{\textrm{Fl}} + k_{\textrm{nr}} + k_{\textrm{qE}} + f_{\textrm{RC}}k_{\textrm{PC}} } \right) \frac{\textrm{N}_{\textrm{ChlA}}}{\textrm{N}_{\textrm{RC}}},
\end{equation}
where $k_{\textrm{PC}}$, $k_{\textrm{fl}}$, $k_{\textrm{nr}}$, and $k_{\textrm{qE}}$ are effective rate constants for photochemistry, fluorescence, non-radiative decay, and qE quenching, respectively. Here, because of perfect competition between open RCs, $\sigma_{\textrm{PC}}$ goes as $\sim$1/$f_{\textrm{RC}}$, as seen in Fig.~3A of the main text. On the other hand, in the puddle model,
\begin{equation}
\label{eq:sigmaP}
\sigma_{\textrm{PC}} = \left( \frac{k_{\textrm{PC}}}{k_{\textrm{Fl}} + k_{\textrm{nr}} + k_{\textrm{qE}} + k_{\textrm{PC}} } \right) \frac{\textrm{N}_{\textrm{ChlA}}}{\textrm{N}_{\textrm{RC}}},
\end{equation}
which means that there is no dependence of $\sigma_{\textrm{PC}}$ on $f_{\textrm{RC}}$, as would be expected in a model without competition between open RCs.

\subsection*{Probability of the quenched conformation: evidence from single molecule data}
\addcontentsline{toc}{subsection}{\protect\numberline{}Probability of the quenched conformation: evidence from single molecule data}%
Kruger, et al., measured the chlorophyll fluorescence of isolated LHCII trimers in a variety of qE mimicking conditions \cite{Kruger2012}. Based on the distribution of chlorophyll fluorescence intensities observed in each condition, the authors assigned a percentage  time the complexes spent in a quenched state. The highest observed such percentage was $\sim$30\%. We have defined the percentage of active quenching sites in our model as the fraction of LHCII monomers with a quenching site ($P_{\textrm{qE}}$). In a quenched trimer, 1, 2, or 3 monomers may be quenched, as LHCII is composed of different homologs \cite{Ballottari2012}. Thus, the upper bound for $P_{\textrm{qE}}$ from the single molecule data is 30\%. The solid grey line in Fig.~4, inset, in the main text indicates the range $P_{\textrm{qE}}<30\%$. Our interpretation is merely a proof of concept of how single molecule data might be related to the quenching measured on intact leaves. A more detailed analysis of the single molecule data is required for a more precise interpretation in the context of our model.

\subsection*{Simple correction to lake/puddle model prediction for photochemical yield}
\begin{figure*}
\includegraphics{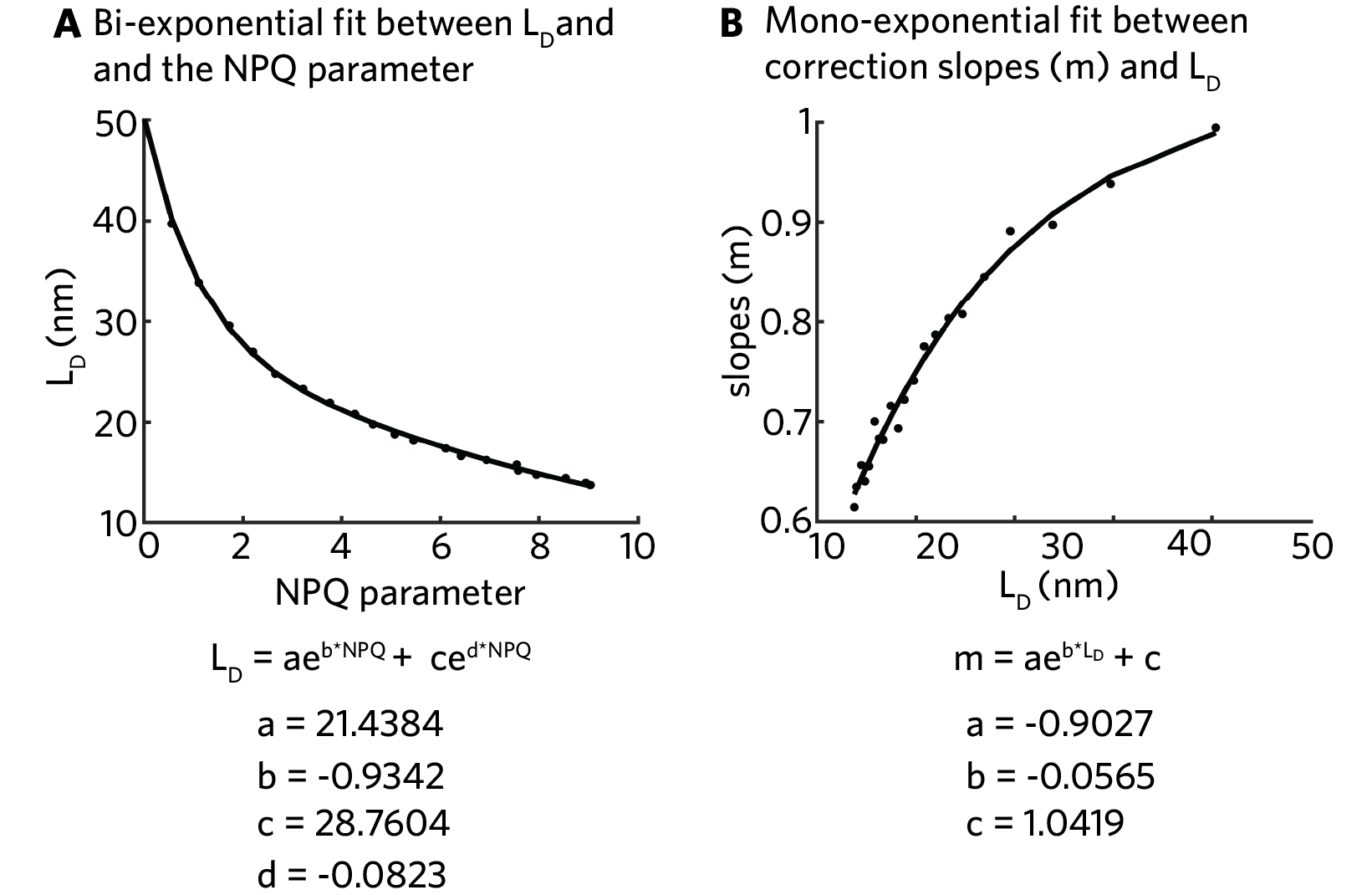}
  \caption{\textbf{Equations for correcting lake/puddle estimation of photochemical yield.} \textbf{(A)} As outlined in the Supplementary Text, a simple correction for $\Phi_{\textrm{PC}}^{\textrm{Fl}}$ ($\Phi_{\textrm{II}}$) begins by measuring the NPQ parameter. The $L_{\textrm{D}}$ can be predicted using the bi-exponential decay indicated, which was fit to the points in the plot. \textbf{(B)} The $L_{\textrm{D}}$ then directly maps to the correction factor ($m$) using the mono-exponential shown, which was fit to the points in the plot. $\Phi_{\textrm{PC}}^{\textrm{dir}}$, the photochemical yield predicted from our model, is simply $m \cdot \Phi_{\textrm{PC}}^{\textrm{Fl}}$. \label{fig:correct}}
  
  \addcontentsline{toc}{subsection}{\protect\numberline{}Fig. S3}%
\end{figure*}

\addcontentsline{toc}{subsection}{\protect\numberline{}Simple correction to lake/puddle model}%
In the main text, we outline a strategy to simply correct the photochemical yield predicted by the lake and puddle models, $\Phi_{\textrm{PC}}^{\textrm{Fl}}$, using only the additional measurement of the NPQ parameter (Eq.~\ref{eq:npq}). We fit a sum of two exponentials to map the one-to-one relationship between the $L_{\textrm{D}}$ and the NPQ parameter (Fig.~\ref{fig:correct}A). As noted in the main text, the slope ($m$) of the line that describes the correspondence between $\Phi_{\textrm{PC}}^{\textrm{Fl}}$ and $\Phi_{\textrm{PC}}^{\textrm{dir}}$, the photochemical yield calculated directly using our model, ($\Phi_{\textrm{PC}}^{\textrm{dir}} = m(L_{\textrm{D}})\cdot \Phi_{\textrm{PC}}^{\textrm{Fl}}$) has a one-to-one correspondence with the $L_{\textrm{D}}$. We fit a mono-exponential rise to map this one-to-one relationship (Fig.~\ref{fig:correct}B). Thus, the measurement of the NPQ parameter gives a measure of the $L_{\textrm{D}}$, which then gives a measure for the correction factor ($m$) for $\Phi_{\textrm{PC}}^{\textrm{Fl}}$. 

The method in steps:
\begin{enumerate}
  \item Measure $\Phi_{\textrm{PC}}^{\textrm{Fl}}$ ($\Phi_{\textrm{II}}$) and NPQ parameter using PAM fluorescence
  \item Calculate $L_{\textrm{D}}$ from NPQ parameter using equation at bottom of Fig.~\ref{fig:correct}A
  \item Calculate $m$ from $L_{\textrm{D}}$ equation at bottom of Fig.~\ref{fig:correct}B
  \item Calculate $\Phi_{\textrm{PC}}^{\textrm{Fl}}$ by multiplying $m$ and $\Phi_{\textrm{PC}}^{\textrm{Fl}}$ 
\end{enumerate} 

It is important to note that the results in Fig.~\ref{fig:correct} represent a numeric interpolation and as such cannot be trusted outside of the range simulated data range (NPQ parameter = 0-9 and \hfill \break $L_{\textrm{D}}$ = 50-15 nm). 

\clearpage

\bibliography{NPQrefs}

\begin{thebibliography}{10}
\expandafter\ifx\csname url\endcsname\relax
  \def\url#1{\texttt{#1}}\fi
\expandafter\ifx\csname urlprefix\endcsname\relax\def\urlprefix{URL }\fi
\providecommand{\bibinfo}[2]{#2}
\providecommand{\eprint}[2][]{\url{#2}}

\bibitem{Field1998}
\bibinfo{author}{Field, C.~B.}, \bibinfo{author}{Behrenfeld, M.~J.},
  \bibinfo{author}{Randerson, J.~T.} \& \bibinfo{author}{Falkowski, P.}
\newblock \bibinfo{title}{Primary production of the biosphere: integrating
  terrestrial and oceanic components}.
\newblock \emph{\bibinfo{journal}{Science}} \textbf{\bibinfo{volume}{281}},
  \bibinfo{pages}{237--240} (\bibinfo{year}{1998}).

\bibitem{Long2015}
\bibinfo{author}{Long, S.~P.}, \bibinfo{author}{Marshall-Colon, A.} \&
  \bibinfo{author}{Zhu, X.-G.}
\newblock \bibinfo{title}{Meeting the global food demand of the future by
  engineering crop photosynthesis and yield potential}.
\newblock \emph{\bibinfo{journal}{Cell}} \textbf{\bibinfo{volume}{161}},
  \bibinfo{pages}{56--66} (\bibinfo{year}{2015}).

\bibitem{Bennett2013}
\bibinfo{author}{Bennett, D. I.~G.}, \bibinfo{author}{Amarnath, K.} \&
  \bibinfo{author}{Fleming, G.~R.}
\newblock \bibinfo{title}{A {Structure}-{Based} {Model} of {Energy} {Transfer}
  {Reveals} the {Principles} of {Light} {Harvesting} in {Photosystem} {II}
  {Supercomplexes}}.
\newblock \emph{\bibinfo{journal}{Journal of the American Chemical Society}}
  \textbf{\bibinfo{volume}{135}}, \bibinfo{pages}{9164--9173}
  (\bibinfo{year}{2013}).

\bibitem{Amarnath2016}
\bibinfo{author}{Amarnath, K.}, \bibinfo{author}{Bennett, D. I.~G.},
  \bibinfo{author}{Schneider, A.~R.} \& \bibinfo{author}{Fleming, G.~R.}
\newblock \bibinfo{title}{Multiscale model of light harvesting by photosystem
  {II} in plants}.
\newblock \emph{\bibinfo{journal}{Proceedings of the National Academy of
  Sciences}} \textbf{\bibinfo{volume}{113}}, \bibinfo{pages}{1156--1161}
  (\bibinfo{year}{2016}).

\bibitem{Kulheim2002}
\bibinfo{author}{K{\"u}lheim, C.}, \bibinfo{author}{{\AA}gren, J.} \&
  \bibinfo{author}{Jansson, S.}
\newblock \bibinfo{title}{Rapid {Regulation} of {Light} {Harvesting} and
  {Plant} {Fitness} in the {Field}}.
\newblock \emph{\bibinfo{journal}{Science}} \textbf{\bibinfo{volume}{297}},
  \bibinfo{pages}{91--93} (\bibinfo{year}{2002}).

\bibitem{Kromdijk2016}
\bibinfo{author}{Kromdijk, J.} \emph{et~al.}
\newblock \bibinfo{title}{Improving photosynthesis and crop productivity by
  accelerating recovery from photoprotection}.
\newblock \emph{\bibinfo{journal}{Science}} \textbf{\bibinfo{volume}{354}},
  \bibinfo{pages}{857--861} (\bibinfo{year}{2016}).

\bibitem{Baker2008}
\bibinfo{author}{Baker, N.~R.}
\newblock \bibinfo{title}{Chlorophyll fluorescence: A probe of photosynthesis
  in vivo}.
\newblock \emph{\bibinfo{journal}{Annual Review of Plant Biology}}
  \textbf{\bibinfo{volume}{59}}, \bibinfo{pages}{89--113}
  (\bibinfo{year}{2008}).

\bibitem{Way2012}
\bibinfo{author}{Way, D.~A.} \& \bibinfo{author}{Pearcy, R.~W.}
\newblock \bibinfo{title}{Sunflecks in trees and forests: from photosynthetic
  physiology to global change biology}.
\newblock \emph{\bibinfo{journal}{Tree Physiology}}
  \textbf{\bibinfo{volume}{32}}, \bibinfo{pages}{1066--1081}
  (\bibinfo{year}{2012}).

\bibitem{Li2002}
\bibinfo{author}{Li, X.-P.}, \bibinfo{author}{M{\"u}ller-Moul{\'e}, P.},
  \bibinfo{author}{Gilmore, A.~M.} \& \bibinfo{author}{Niyogi, K.~K.}
\newblock \bibinfo{title}{Psbs-dependent enhancement of feedback de-excitation
  protects photosystem {II} from photoinhibition}.
\newblock \emph{\bibinfo{journal}{Proceedings of the National Academy of
  Sciences}} \textbf{\bibinfo{volume}{99}}, \bibinfo{pages}{15222--15227}
  (\bibinfo{year}{2002}).

\bibitem{Genty1989}
\bibinfo{author}{Genty, B.}, \bibinfo{author}{Briantais, J.-M.} \&
  \bibinfo{author}{Baker, N.~R.}
\newblock \bibinfo{title}{The relationship between the quantum yield of
  photosynthetic electron transport and quenching of chlorophyll fluorescence}.
\newblock \emph{\bibinfo{journal}{Biochimica et Biophysica Acta (BBA)-General
  Subjects}} \textbf{\bibinfo{volume}{990}}, \bibinfo{pages}{87--92}
  (\bibinfo{year}{1989}).

\bibitem{Kanazawa2002}
\bibinfo{author}{Kanazawa, A.} \& \bibinfo{author}{Kramer, D.~M.}
\newblock \bibinfo{title}{In vivo modulation of nonphotochemical exciton
  quenching (npq) by regulation of the chloroplast atp synthase}.
\newblock \emph{\bibinfo{journal}{Proceedings of the National Academy of
  Sciences}} \textbf{\bibinfo{volume}{99}}, \bibinfo{pages}{12789--12794}
  (\bibinfo{year}{2002}).

\bibitem{Duffy2015}
\bibinfo{author}{Duffy, C. D.~P.} \& \bibinfo{author}{Ruban, A.~V.}
\newblock \bibinfo{title}{Dissipative pathways in the photosystem-{II} antenna
  in plants}.
\newblock \emph{\bibinfo{journal}{Journal of Photochemistry and Photobiology B:
  Biology}} \textbf{\bibinfo{volume}{152}}, \bibinfo{pages}{215--226}
  (\bibinfo{year}{2015}).

\bibitem{Robinson1966}
\bibinfo{author}{Robinson, G.~W.}
\newblock \bibinfo{title}{Excitation transfer and trapping in photosynthesis.}
\newblock In \emph{\bibinfo{booktitle}{Brookhaven Symposia in Biology}},
  vol.~\bibinfo{volume}{19}, \bibinfo{pages}{16} (\bibinfo{year}{1966}).

\bibitem{Kramer2004}
\bibinfo{author}{Kramer, D.~M.}, \bibinfo{author}{Johnson, G.},
  \bibinfo{author}{Kiirats, O.} \& \bibinfo{author}{Edwards, G.~E.}
\newblock \bibinfo{title}{New fluorescence parameters for the determination of
  q$_a$ redox state and excitation energy fluxes}.
\newblock \emph{\bibinfo{journal}{Photosynthesis Research}}
  \textbf{\bibinfo{volume}{79}}, \bibinfo{pages}{209--218}
  (\bibinfo{year}{2004}).

\bibitem{Caffarri2009}
\bibinfo{author}{Caffarri, S.}, \bibinfo{author}{Kou\v{r}il, R.},
  \bibinfo{author}{Kereiche, S.}, \bibinfo{author}{Boekema, E.~J.} \&
  \bibinfo{author}{Croce, R.}
\newblock \bibinfo{title}{Functional architecture of higher plant photosystem
  {II} supercomplexes}.
\newblock \emph{\bibinfo{journal}{EMBO J.}} \textbf{\bibinfo{volume}{28}},
  \bibinfo{pages}{3052--3063} (\bibinfo{year}{2009}).

\bibitem{Liu2004}
\bibinfo{author}{Liu, Z.} \emph{et~al.}
\newblock \bibinfo{title}{Crystal structure of spinach major light-harvesting
  complex at 2.72 {{\AA}} resolution}.
\newblock \emph{\bibinfo{journal}{Nature}} \textbf{\bibinfo{volume}{428}},
  \bibinfo{pages}{287--292} (\bibinfo{year}{2004}).

\bibitem{Ishizaki2010}
\bibinfo{author}{Ishizaki, A.}, \bibinfo{author}{Calhoun, T.~R.},
  \bibinfo{author}{Schlau-Cohen, G.~S.} \& \bibinfo{author}{Fleming, G.~R.}
\newblock \bibinfo{title}{Quantum coherence and its interplay with protein
  environments in photosynthetic electronic energy transfer}.
\newblock \emph{\bibinfo{journal}{Physical Chemistry Chemical Physics}}
  \textbf{\bibinfo{volume}{12}}, \bibinfo{pages}{7319--7337}
  (\bibinfo{year}{2010}).

\bibitem{Sumi1999}
\bibinfo{author}{Sumi, H.}
\newblock \bibinfo{title}{Theory on rates of excitation-energy transfer between
  molecular aggregates through distributed transition dipoles with application
  to the antenna system in bacterial photosynthesis}.
\newblock \emph{\bibinfo{journal}{The Journal of Physical Chemistry B}}
  \textbf{\bibinfo{volume}{103}}, \bibinfo{pages}{252--260}
  (\bibinfo{year}{1999}).

\bibitem{Scholes2000}
\bibinfo{author}{Scholes, G.~D.} \& \bibinfo{author}{Fleming, G.~R.}
\newblock \bibinfo{title}{On the mechanism of light harvesting in
  photosynthetic purple bacteria: {B800} to {B850} energy transfer}.
\newblock \emph{\bibinfo{journal}{The Journal of Physical Chemistry B}}
  \textbf{\bibinfo{volume}{104}}, \bibinfo{pages}{1854--1868}
  (\bibinfo{year}{2000}).

\bibitem{Roden2016}
\bibinfo{author}{Roden, J. J.~J.}, \bibinfo{author}{Bennett, D. I.~G.} \&
  \bibinfo{author}{Whaley, K.~B.}
\newblock \bibinfo{title}{Long-range energy transport in photosystem {II}}.
\newblock \emph{\bibinfo{journal}{The Journal of Chemical Physics}}
  \textbf{\bibinfo{volume}{144}}, \bibinfo{pages}{245101}
  (\bibinfo{year}{2016}).

\bibitem{Kreisbeck2014}
\bibinfo{author}{Kreisbeck, C.}, \bibinfo{author}{Kramer, T.} \&
  \bibinfo{author}{Aspuru-Guzik, A.}
\newblock \bibinfo{title}{Scalable high-performance algorithm for the
  simulation of exciton dynamics. application to the light-harvesting complex
  {II} in the presence of resonant vibrational modes}.
\newblock \emph{\bibinfo{journal}{Journal of Chemical Theory and Computation}}
  \textbf{\bibinfo{volume}{10}}, \bibinfo{pages}{4045--4054}
  (\bibinfo{year}{2014}).

\bibitem{Schneider2013}
\bibinfo{author}{Schneider, A.} \& \bibinfo{author}{Geissler, P.}
\newblock \bibinfo{title}{Coexistence of {Fluid} and {Crystalline} {Phases} of
  {Proteins} in {Photosynthetic} {Membranes}}.
\newblock \emph{\bibinfo{journal}{Biophysical Journal}}
  \textbf{\bibinfo{volume}{105}}, \bibinfo{pages}{1161--1170}
  (\bibinfo{year}{2013}).

\bibitem{Joliot1964}
\bibinfo{author}{Joliot, A.} \& \bibinfo{author}{Joliot, P.}
\newblock \bibinfo{title}{Etude cinetique de la reaction photochimique liberant
  loxygene au cours de la photosynthese}.
\newblock \emph{\bibinfo{journal}{Comptes Rendus Hebdomadaires Des Seances De L
  Academie Des Sciences}} \textbf{\bibinfo{volume}{258}}, \bibinfo{pages}{4622}
  (\bibinfo{year}{1964}).

\bibitem{Chmeliov2016a}
\bibinfo{author}{Chmeliov, J.}, \bibinfo{author}{Trinkunas, G.},
  \bibinfo{author}{Amerongen, H.} \& \bibinfo{author}{Valkunas, L.}
\newblock \bibinfo{title}{Excitation migration in fluctuating light-harvesting
  antenna systems.}
\newblock \emph{\bibinfo{journal}{Photosynthesis Research}}
  \textbf{\bibinfo{volume}{127}} (\bibinfo{year}{2016}).

\bibitem{vanOort2010}
\bibinfo{author}{van Oort, B.} \emph{et~al.}
\newblock \bibinfo{title}{Effect of antenna-depletion in photosystem {II} on
  excitation energy transfer in arabidopsis thaliana}.
\newblock \emph{\bibinfo{journal}{Biophysical Journal}}
  \textbf{\bibinfo{volume}{98}}, \bibinfo{pages}{922--931}
  (\bibinfo{year}{2010}).

\bibitem{Sylak-Glassman2016}
\bibinfo{author}{Sylak-Glassman, E.~J.}, \bibinfo{author}{Zaks, J.},
  \bibinfo{author}{Amarnath, K.}, \bibinfo{author}{Leuenberger, M.} \&
  \bibinfo{author}{Fleming, G.~R.}
\newblock \bibinfo{title}{Characterizing non-photochemical quenching in leaves
  through fluorescence lifetime snapshots}.
\newblock \emph{\bibinfo{journal}{Photosynthesis Research}}
  \textbf{\bibinfo{volume}{127}}, \bibinfo{pages}{69--76}
  (\bibinfo{year}{2016}).

\bibitem{Gruber2016}
\bibinfo{author}{Gruber, J.~M.} \emph{et~al.}
\newblock \bibinfo{title}{Dynamic quenching in single photosystem {II}
  supercomplexes}.
\newblock \emph{\bibinfo{journal}{Physical Chemistry Chemical Physics}}
  \textbf{\bibinfo{volume}{18}}, \bibinfo{pages}{25852--25860}
  (\bibinfo{year}{2016}).

\bibitem{Ruban2007}
\bibinfo{author}{Ruban, A.~V.} \emph{et~al.}
\newblock \bibinfo{title}{Identification of a mechanism of photoprotective
  energy dissipation in higher plants}.
\newblock \emph{\bibinfo{journal}{Nature}} \textbf{\bibinfo{volume}{450}},
  \bibinfo{pages}{575--578} (\bibinfo{year}{2007}).

\bibitem{Kruger2012}
\bibinfo{author}{Kr{\"u}ger, T.~J.} \emph{et~al.}
\newblock \bibinfo{title}{Controlled {Disorder} in {Plant} {Light}-{Harvesting}
  {Complex} {II} {Explains} {Its} {Photoprotective} {Role}}.
\newblock \emph{\bibinfo{journal}{Biophysical Journal}}
  \textbf{\bibinfo{volume}{102}}, \bibinfo{pages}{2669--2676}
  (\bibinfo{year}{2012}).

\bibitem{Duffy2012}
\bibinfo{author}{Duffy, C. D.~P.} \emph{et~al.}
\newblock \bibinfo{title}{Modeling of {Fluorescence} {Quenching} by {Lutein} in
  the {Plant} {Light}-{Harvesting} {Complex} {LHCII}}.
\newblock \emph{\bibinfo{journal}{The Journal of Physical Chemistry B}}
  \textbf{\bibinfo{volume}{117}}, \bibinfo{pages}{10974--10986}
  (\bibinfo{year}{2012}).

\bibitem{Ahn2008}
\bibinfo{author}{Ahn, T.~K.} \emph{et~al.}
\newblock \bibinfo{title}{Architecture of a {Charge}-{Transfer} {State}
  {Regulating} {Light} {Harvesting} in a {Plant} {Antenna} {Protein}}.
\newblock \emph{\bibinfo{journal}{Science}} \textbf{\bibinfo{volume}{320}},
  \bibinfo{pages}{794--797} (\bibinfo{year}{2008}).

\bibitem{Avenson2009}
\bibinfo{author}{Avenson, T.~J.} \emph{et~al.}
\newblock \bibinfo{title}{Lutein can act as a switchable charge transfer
  quencher in the {CP26} light-harvesting complex}.
\newblock \emph{\bibinfo{journal}{Journal of Biological Chemistry}}
  \textbf{\bibinfo{volume}{284}}, \bibinfo{pages}{2830--2835}
  (\bibinfo{year}{2009}).

\bibitem{Schlau-Cohen2015}
\bibinfo{author}{Schlau-Cohen, G.~S.} \emph{et~al.}
\newblock \bibinfo{title}{Single-{Molecule} {Identification} of {Quenched} and
  {Unquenched} {States} of {LHCII}}.
\newblock \emph{\bibinfo{journal}{The Journal of Physical Chemistry Letters}}
  \textbf{\bibinfo{volume}{6}}, \bibinfo{pages}{860--867}
  (\bibinfo{year}{2015}).

\bibitem{Niyogi2001}
\bibinfo{author}{Niyogi, K.~K.} \emph{et~al.}
\newblock \bibinfo{title}{Photoprotection in a zeaxanthin-and lutein-deficient
  double mutant of \emph{{A}rabidopsis}}.
\newblock \emph{\bibinfo{journal}{Photosynthesis Research}}
  \textbf{\bibinfo{volume}{67}}, \bibinfo{pages}{139--145}
  (\bibinfo{year}{2001}).

\bibitem{Zaks2012}
\bibinfo{author}{Zaks, J.}, \bibinfo{author}{Amarnath, K.},
  \bibinfo{author}{Kramer, D.~M.}, \bibinfo{author}{Niyogi, K.~K.} \&
  \bibinfo{author}{Fleming, G.~R.}
\newblock \bibinfo{title}{A kinetic model of rapidly reversible
  nonphotochemical quenching}.
\newblock \emph{\bibinfo{journal}{Proceedings of the National Academy of
  Sciences}} \textbf{\bibinfo{volume}{109}}, \bibinfo{pages}{15757--15762}
  (\bibinfo{year}{2012}).

\bibitem{Porcar-Castell2014}
\bibinfo{author}{Porcar-Castell, A.} \emph{et~al.}
\newblock \bibinfo{title}{Linking chlorophyll a fluorescence to photosynthesis
  for remote sensing applications: mechanisms and challenges}.
\newblock \emph{\bibinfo{journal}{Journal of Experimental Botany}}
  \textbf{\bibinfo{volume}{65}}, \bibinfo{pages}{4065--4095}
  (\bibinfo{year}{2014}).

\bibitem{Zhu2013}
\bibinfo{author}{Zhu, X.-G.}, \bibinfo{author}{Wang, Y.}, \bibinfo{author}{Ort,
  D.~R.} \& \bibinfo{author}{Long, S.~P.}
\newblock \bibinfo{title}{e-photosynthesis: a comprehensive dynamic mechanistic
  model of c3 photosynthesis: from light capture to sucrose synthesis}.
\newblock \emph{\bibinfo{journal}{Plant, Cell \& Environment}}
  \textbf{\bibinfo{volume}{36}}, \bibinfo{pages}{1711--1727}
  (\bibinfo{year}{2013}).

\bibitem{Lazar1999}
\bibinfo{author}{Laz{\'a}r, D.}
\newblock \bibinfo{title}{Chlorophyll a fluorescence induction}.
\newblock \emph{\bibinfo{journal}{Biochimica et Biophysica Acta
  (BBA)-Bioenergetics}} \textbf{\bibinfo{volume}{1412}}, \bibinfo{pages}{1--28}
  (\bibinfo{year}{1999}).

\bibitem{Wan2015}
\bibinfo{author}{Wan, Y.} \emph{et~al.}
\newblock \bibinfo{title}{Cooperative singlet and triplet exciton transport in
  tetracene crystals visualized by ultrafast microscopy}.
\newblock \emph{\bibinfo{journal}{Nature Chemistry}}
  \textbf{\bibinfo{volume}{7}}, \bibinfo{pages}{785--792}
  (\bibinfo{year}{2015}).

\bibitem{Penwell2017}
\bibinfo{author}{Penwell, S.~B.}, \bibinfo{author}{Ginsberg, L.~D.},
  \bibinfo{author}{Noriega, R.} \& \bibinfo{author}{Ginsberg, N.~S.}
\newblock \bibinfo{title}{Resolving ultrafast exciton migration in organic
  solids at the nanoscale}.
\newblock \emph{\bibinfo{journal}{Nature Materials}}
  \textbf{\bibinfo{volume}{16}}, \bibinfo{pages}{1136} (\bibinfo{year}{2017}).

\bibitem{Stirbet2011}
\bibinfo{author}{Stirbet, A.} \& \bibinfo{author}{Govindjee}.
\newblock \bibinfo{title}{On the relation between the {K}autsky effect
  (chlorophyll a fluorescence induction) and photosystem {II}: basics and
  applications of the {OJIP} fluorescence transient}.
\newblock \emph{\bibinfo{journal}{Journal of Photochemistry and Photobiology B:
  Biology}} \textbf{\bibinfo{volume}{104}}, \bibinfo{pages}{236--257}
  (\bibinfo{year}{2011}).

\bibitem{DallOsto2017}
\bibinfo{author}{Dall'Osto, L.} \emph{et~al.}
\newblock \bibinfo{title}{Two mechanisms for dissipation of excess light in
  monomeric and trimeric light-harvesting complexes.}
\newblock \emph{\bibinfo{journal}{Nature Plants}} \textbf{\bibinfo{volume}{3}},
  \bibinfo{pages}{17033} (\bibinfo{year}{2017}).

\bibitem{Zaks2013}
\bibinfo{author}{Zaks, J.}, \bibinfo{author}{Amarnath, K.},
  \bibinfo{author}{Sylak-Glassman, E.~J.} \& \bibinfo{author}{Fleming, G.~R.}
\newblock \bibinfo{title}{Models and measurements of energy-dependent
  quenching}.
\newblock \emph{\bibinfo{journal}{Photosynthesis Research}}
  \textbf{\bibinfo{volume}{116}}, \bibinfo{pages}{389--409}
  (\bibinfo{year}{2013}).

\bibitem{Kouril2012}
\bibinfo{author}{Kou\v{r}il, R.}, \bibinfo{author}{Dekker, J.~P.} \&
  \bibinfo{author}{Boekema, E.~J.}
\newblock \bibinfo{title}{Supramolecular organization of photosystem {II} in
  green plants}.
\newblock \emph{\bibinfo{journal}{Biochimica et Biophysica Acta ({BBA}) -
  Bioenergetics}} \textbf{\bibinfo{volume}{1817}}, \bibinfo{pages}{2--12}
  (\bibinfo{year}{2012}).

\bibitem{Raszewski2008}
\bibinfo{author}{Raszewski, G.} \& \bibinfo{author}{Renger, T.}
\newblock \bibinfo{title}{Light harvesting in photosystem {II} core complexes
  is limited by the transfer to the trap: can the core complex turn into a
  photoprotective mode?}
\newblock \emph{\bibinfo{journal}{Journal of the American Chemical Society}}
  \textbf{\bibinfo{volume}{130}}, \bibinfo{pages}{4431--4446}
  (\bibinfo{year}{2008}).

\bibitem{Novoderezhkin2011a}
\bibinfo{author}{Novoderezhkin, V.}, \bibinfo{author}{Marin, A.} \&
  \bibinfo{author}{van Grondelle, R.}
\newblock \bibinfo{title}{Intra-and inter-monomeric transfers in the light
  harvesting {LHCII} complex: the {R}edfield--{F}{\"o}rster picture}.
\newblock \emph{\bibinfo{journal}{Physical Chemistry Chemical Physics}}
  \textbf{\bibinfo{volume}{13}}, \bibinfo{pages}{17093--17103}
  (\bibinfo{year}{2011}).

\bibitem{Caffarri2011}
\bibinfo{author}{Caffarri, S.}, \bibinfo{author}{Broess, K.},
  \bibinfo{author}{Croce, R.} \& \bibinfo{author}{van Amerongen, H.}
\newblock \bibinfo{title}{Excitation {Energy} {Transfer} and {Trapping} in
  {Higher} {Plant} {Photosystem} {II} {Complexes} with {Different} {Antenna}
  {Sizes}}.
\newblock \emph{\bibinfo{journal}{Biophysical Journal}}
  \textbf{\bibinfo{volume}{100}}, \bibinfo{pages}{2094--2103}
  (\bibinfo{year}{2011}).

\bibitem{Miloslavina2008}
\bibinfo{author}{Miloslavina, Y.} \emph{et~al.}
\newblock \bibinfo{title}{Far-red fluorescence: A direct spectroscopic marker
  for {LHCII} oligomer formation in non-photochemical quenching}.
\newblock \emph{\bibinfo{journal}{FEBS letters}}
  \textbf{\bibinfo{volume}{582}}, \bibinfo{pages}{3625--3631}
  (\bibinfo{year}{2008}).

\bibitem{Holt2005}
\bibinfo{author}{Holt, N.~E.} \emph{et~al.}
\newblock \bibinfo{title}{Carotenoid cation formation and the regulation of
  photosynthetic light harvesting}.
\newblock \emph{\bibinfo{journal}{Science}} \textbf{\bibinfo{volume}{307}},
  \bibinfo{pages}{433--436} (\bibinfo{year}{2005}).

\bibitem{Ruban2012a}
\bibinfo{author}{Ruban, A.~V.}, \bibinfo{author}{Johnson, M.~P.} \&
  \bibinfo{author}{Duffy, C.~D.}
\newblock \bibinfo{title}{The photoprotective molecular switch in the
  photosystem {II} antenna}.
\newblock \emph{\bibinfo{journal}{Biochimica et Biophysica Acta
  (BBA)-Bioenergetics}} \textbf{\bibinfo{volume}{1817}},
  \bibinfo{pages}{167--181} (\bibinfo{year}{2012}).

\bibitem{Betterle2009}
\bibinfo{author}{Betterle, N.} \emph{et~al.}
\newblock \bibinfo{title}{Light-induced dissociation of an antenna
  hetero-oligomer is needed for non-photochemical quenching induction}.
\newblock \emph{\bibinfo{journal}{Journal of Biological Chemistry}}
  \textbf{\bibinfo{volume}{284}}, \bibinfo{pages}{15255--15266}
  (\bibinfo{year}{2009}).

\bibitem{Johnson2011}
\bibinfo{author}{Johnson, M.~P.} \emph{et~al.}
\newblock \bibinfo{title}{Photoprotective {Energy} {Dissipation} {Involves} the
  {Reorganization} of {Photosystem} {II} {Light}-{Harvesting} {Complexes} in
  the {Grana} {Membranes} of {Spinach} {Chloroplasts}}.
\newblock \emph{\bibinfo{journal}{The Plant Cell}}
  \textbf{\bibinfo{volume}{23}}, \bibinfo{pages}{1468--1479}
  (\bibinfo{year}{2011}).

\bibitem{Ahn2009}
\bibinfo{author}{Ahn, T.~K.} \emph{et~al.}
\newblock \bibinfo{title}{Investigating energy partitioning during
  photosynthesis using an expanded quantum yield convention}.
\newblock \emph{\bibinfo{journal}{Chemical Physics}}
  \textbf{\bibinfo{volume}{357}}, \bibinfo{pages}{151--158}
  (\bibinfo{year}{2009}).

\bibitem{Ihalainen2005}
\bibinfo{author}{Ihalainen, J.~A.} \emph{et~al.}
\newblock \bibinfo{title}{Kinetics of excitation trapping in intact photosystem
  {I} of {C}hlamydomonas reinhardtii and {A}rabidopsis thaliana}.
\newblock \emph{\bibinfo{journal}{Biochimica et Biophysica Acta (BBA) -
  Bioenergetics}} \textbf{\bibinfo{volume}{1706}}, \bibinfo{pages}{267--275}
  (\bibinfo{year}{2005}).

\bibitem{Malkin1981}
\bibinfo{author}{Malkin, S.}, \bibinfo{author}{Armond, P.~A.},
  \bibinfo{author}{Mooney, H.~A.} \& \bibinfo{author}{Fork, D.~C.}
\newblock \bibinfo{title}{Photosystem {II} photosynthetic unit sizes from
  fluorescence induction in leaves {CORRELATION TO PHOTOSYNTHETIC CAPACITY}}.
\newblock \emph{\bibinfo{journal}{Plant Physiology}}
  \textbf{\bibinfo{volume}{67}}, \bibinfo{pages}{570--579}
  (\bibinfo{year}{1981}).

\bibitem{Ballottari2012}
\bibinfo{author}{Ballottari, M.}, \bibinfo{author}{Girardon, J.},
  \bibinfo{author}{Dall'Osto, L.} \& \bibinfo{author}{Bassi, R.}
\newblock \bibinfo{title}{Evolution and functional properties of photosystem
  {II} light harvesting complexes in eukaryotes}.
\newblock \emph{\bibinfo{journal}{Biochimica et Biophysica Acta
  (BBA)-Bioenergetics}} \textbf{\bibinfo{volume}{1817}},
  \bibinfo{pages}{143--157} (\bibinfo{year}{2012}).

\end{thebibliography}
\bibliographystyle{naturemag}

\end{document}